# 7-Methylquinolinium Iodobismuthate Memristor: Exploring Plasticity and Memristive Properties for Digit Classification in Physical Reservoir Computing

Gisya Abdi,[1]* Ahmet Karacali,[2] Alif Syafiq Kamarol Zaman,[2] Marlena Gryl,[3]* Andrzej Sławek,[1] Aleksandra Szkudlarek,[1] Hirofumi Tanaka,[2,4] Konrad Szaciłowski[1,5]*

[1]Academic Centre for Materials and Nanotechnology, AGH University of Krakow, Kawiory 30, 30-055 Kraków, Poland

[2]Graduate School of Life Science and Systems Engineering, Kyushu Institute of Technology (Kyutech)

[3]Faculty of Chemistry, Jagiellonian University, Gronostajowa 2, 30-387 Kraków, Poland

[4]Research Center for Neuromorphic AI Hardware, Kyushu Institute of Technology (Kyutech), Kitakyushu 808-0196, Japan

[5]Unconventional Computing Lab, University of the West of England, Bristol BS16 1QY, United Kingdom

*Corresponding authors: agisya@agh.edu.pl, szacilow@agh.edu.pl, marlena.gryl@uj.edu.pl

**Abstract.** This study investigates 7-methylquinolinium halobismuthates (I, Br, and Cl) in two aspects: (1) their structural and semiconducting properties influenced by anionic composition, and (2) their memristive and plasticity characteristics for neuromorphic and reservoir computing applications. Structural changes induced by halides form low-dimensional halobismuthate fragments, confirmed by crystallographic analysis. Optical band gaps were studied using diffuse reflectance spectroscopy, aligning with density functional theory results. Due to solubility limitations, only bismuth iodide complexes were explored in electronic devices. Current-voltage scans showed pinched hysteresis loops, characteristic of memristors. Conductivity versus temperature study indicates combined ionic and electronic contributions to conductivity of the devices. Given that a memristor can function as a single synapse without the need for programming, aligning with the requirements of neuromorphic computing, the study investigated long-term depression, potentiation, and spike-time-dependent plasticity. As the potentiation-depression plots showed non-linearity with fading memory, these materials can be a good candidate for application in physical reservoir computing. To further assess this material, an electronic device with sixteen gold electrodes was applied, featuring one input and 15 output electrodes deposited on silicon substrate and covered with a layer of studied compound. Basic test to assess the complexity and non-linearity of the devices were conducted through a series of benchmark tasks, including waveform generation, NARMA-2, memory capacity assessment, and noise study under both DC and AC current. The ability of device in MNIST digit classification with 82.26% accuracy and voice classification for digit 2 for six different people with 82 % accuracy has been demonstrated.

1. Introduction

Memristors (or memristive devices), initially considered as a new type of solid-state memory, became the most promising candidates for neuromorphic unconventional computing, especially for computational in material approaches related to reservoir computing [1, 2]. Researchers worldwide are actively investigating diverse materials and configurations for various information processing applications [3-5], aiming to enhance their applicability in nonvolatile memory, logic circuits, and neuromorphic computing [6]. The investigation of materials for resistive switching (RS) devices has encompassed a broad spectrum of compounds [7, 8], including metal oxides [8], metal halides [9], chalcogenides [10], organic- [11] and metal-organic frameworks [12], biomaterials [13] 2D nanomaterials [14] and carbon-based nanomaterials [15]. These materials are of significant interest due to their distinct physicochemical properties, which enable their integration into advanced memory technologies and neuromorphic computing systems. Lead- and bismuth-based perovskites as metal halides hold significant potential for advanced applications beyond photovoltaics, including photodetectors, light-emitting devices, and various other optoelectronic technologies [16, 17]. Although lead halide perovskites are renowned for their high photoconversion efficiency in solar cells, concerns regarding their toxicity and as well as poor stability against environmental factors have prompted the search for safer and more stable alternatives. In contrast, low-dimensional metal halide perovskites have garnered attention as promising alternatives to traditional three-dimensional (3D) structures, especially due to increased stability against hydrolysis and tunability of their properties due to the presence of quantum size effect [18]. Attempts to replace Pb(II) with Sn(II), Ge(II) or their combinations have resulted in compounds that quickly oxidize to Sn(IV) and Ge(IV) upon exposure to air [19, 20]. Bismuth halide complexes present an environmentally friendly option with good chemical stability and tunable band gaps and performance enhancement through composition changes [21]. In contrast to Pb(II) and Sn(II), analogous compounds based on $Bi^{3+}$ ions do not form typical 3D perovskite structures like but instead create low-dimensional assemblies, including 0D octahedral units, 1D chains, and 2D layers by sharing edges, vertices or faces of iodobismuthate octahedra [22-25]. The diverse chemistry and structural versatility of these hybrid halobismuthate salts arise from the selection of various organic cations and halides [26, 27]. These can be influenced by stoichiometry, reaction conditions and the size, shape and dipole moment of counterions [28]. Bismuth complexes have already demonstrated their applicability in various neuromorphic applications, among others in artificial synaptic devices [28-31]. Up to date, they have not been used to perform more advanced computational tasks, e.g. reservoir computing, however their potential in feed-forward neural networks has been demonstrated [31].

At the same time search for new computational paradigms gave rise to reservoir computing, an unconventional scheme of information processing which explores complex dynamics, chaos and random connectivity as computational media [32-36]. Physical reservoir computing is an emerging concept in unconventional computing, focusing on implementing reservoir computing concepts: information processing in disordered systems and media on the basis of their dynamics and evolution [37, 38], in physical systems [39-41]. Unlike artificial neural networks, which struggle with precise connection adjustments and complex training protocols, reservoir computing, and random neural networks offer a feasible alternative [42]. They enable power-efficient, in-memory computing in random networks without the need for precise adjustment of synaptic connections. Reservoir computing, as a computational framework, employs a dynamic reservoir, which usually is a complex nonlinear system, as a key component for processing information [43, 44]. The unique electronic and optical properties of low-dimensional perovskites complexes as memristors significantly enhance computational efficiency in neuromorphic computing [45, 46]. Bismuth complexes offer tunable band gaps, mixed ionic-electronic conductivity and nonlinear current-voltage characteristics along with significant hysteresis make them promising materials for unconventional computing systems. Additionally, we believe that the stability and ease of synthesis of bismuth complexes renders them appealing candidates for practical physical reservoir computing applications. An efficient physical reservoir are vital in computing due to their ability to exploit the complex dynamical behaviors of physical systems, transforming them into powerful tools for solving intricate computational problems [47, 48]. Furthermore, researchers are increasingly leaning toward physical reservoir computing over conventional von Neumann architectures due to possible enhancement of their efficiency (especially in complex tasks of pattern recognition) and reduced energy consumption [49].

## 2. Experimental section

### 2.1 Synthesis of pyridinium-based bismuth iodides

The compounds were produced through the reaction outlined in Scheme 1. Initially, a mixture of 2 mmol $BiI_3$ (or $BiBr_3$, $BiCl_3$) and HI (or HBr, HCl) was added to a hydrothermal reactor followed by sequential addition of 3mmol of protonated 7-methylquinoline (7-Meq) with the corresponding hydrohalic acids (Scheme 1). The vessels were placed in an oven at 100 °C for 5 h, then gradually cooled over a period of 12 h. The resulting crystals of 7-MeqBiI$_3$, 7-MeqBiBr$_3$, 7-MeqBiCl$_3$, were washed with diethyl ether and dried under vacuum. The X-ray diffraction data for the corresponding crystal structures are reported in Table S1.

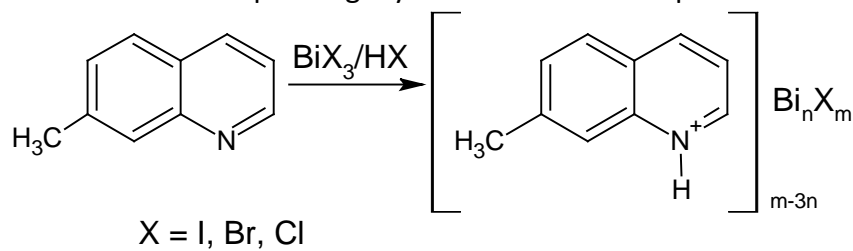

X = I, Br, Cl

Scheme 1.

### 2.2 Materials and characterisation techniques

Single crystal X-ray diffraction experiments were carried out on an XtaLAB Synergy-S X-ray diffractometer using MoK radiation at 100-130K. Data collection and reduction were performed using CrysalisPro 1.171.41.122a (Rigaku OD, 2021) software. First, the structures were solved with direct methods [50]. Then, refinement was carried out with Shelxl [51] as incorporated in a WinGX package [52]. Diffuse reflectance spectra of powder samples were recorded on LAMBDA 750 UV/vis/NIR spectrophotometer (PerkinElmer Inc., USA) in $BaSO_4$ matrix. The same material has been used as a reference. Absorption spectra of studied materials in solution were recorded on a single beam diode-array spectrophotometer (model 8453, Agilent, USA) in quartz cells of 1 cm optical path length. SEM analysis were performed with beam acceleration voltage 15kV, beam current 4 nA, imaging mode secondary electrons, working distance 10 mm. Cross-sectional SEM was measured by FEI Versa 3D double-beam high-resolution scanning electron microscope with beam acceleration voltage 30kV, beam current 0.55 pA, imaging mode secondary electrons, working distance 2.2 mm. beam acceleration voltage 15kV, beam current 4 nA, imaging mode secondary electrons, working distance 10 mm. EDS analysis was performed with dwell time per pixel 2 ms, resolution 512×352 pixels.

The thickness of the layers was estimated by DektakXT stylus profilometer (Bruker DektakXT). The Spectroscopic Ellipsometer SENresearch 4.0 model 850 (SENTECH Instruments GmbH, Germany) was utilized to assess the thickness and optical properties of the semiconductor layers. Thermal evaporation was applied for physical vapor deposition of the metal electrode under vacuum ($9\times10^{-5}$). The Polos SPIN150i spin coater was applied for the thin layer fabrication process. I-V plot measurements, on-off ratio, memristor state retention, and various plasticity measurements were conducted using an SP-150 potentiostate (BioLogic, France) in a two-electrode configuration. The working electrode (WE) was connected to the top metal electrodes on thin layer, while counter electrode (CE) and reference electrode (RE) were grounded and connected to the ITO layer. Spike timing-dependent plasticity (STDP) were investigated using waveform generator connected to analog control input of Autolab PGSTAT 302N potentiostat (Metrohm AG, Switzerland). The benchmark tests for physical reservoir was performed on a 16-electrode setup of $SiO_2$/Si as the substrate with Al as the electrodes and the material was drop-casted on the 100 μm gap. The measurement setup includes a DAQ SCB-68A connected to NI PXIe-1078 chassis (National Instruments). For waveform generation, NARMA2 and MC tasks, a series of sine wave (5 V), uniform white noise (peak-to-peak amplitude of ca. 10 mV and 1000 rate) and random rectangular pulses with 5V amplitude different range of width for 60 s with 1000 points/second rate were given respectively as the input to the device and the 15 were collected for analyses and training. To test performance of physical reservoir device, voice recognition and benchmark tasks were performed by using NI-MultiDAQ which feeds one input and records 15 output simultaneously.

Band structures and density of states were calculated using density functional theory (DFT) with the CASTEP software [53, 54]. The experimental crystal structures were optimised with the BFGS minimization scheme with the generalized gradient approximation (GGA) using the Perdew-Burker-Ernzerhof (PBE) exchange-correlation functional [55]. Van der Waals interactions were included using semi-empirical dispersion correction (DFT-D) with the Tkatchenko-Scheffler (TS) scheme for the exchange functional corrections. CASTEP-specific on-the-fly generated (OTFG) ultrasoft pseudopotentials were used to describe the electron-ion interactions. The convergence criteria for geometry optimization were set to $10^{-5}$ eV/atom for total energy, 0.03 eV/Å for forces, 0.05 GPa for stress, and $10^{-3}$ Å for ionic displacements. For the electron spectroscopy, the convergence criterion was set to $10^{-5}$ eV/atom. The Brillouin zone was sampled with 0.005 Å$^{-1}$ k-point spacing. Gaussian-like Fermi smearing was used.

### 3. Results and discussion
*3.1 Characterization of the bismuth complexes*

Composition and temperature are two key factors influencing the growth of ionic fragments in low-dimensional crystal structures, as demonstrated by the two reported structures containing 0D anionic [Bi$_2$X$_{10}$] units (where X = Br, I), while the chlorobismuthate structures are composed of inorganic chains, [BiCl$_5$]$_n$ (see Figures S1-S3). Structure of 7-MeqBiBr$_3$ belongs to the triclinic crystal system, whereas 7-MeqBiI$_3$ and 7-MeqBiCl$_3$ follow the symmetry of the monoclinic C2/c space groups, respectively. Structural components with a numbering scheme are presented in Figure 1.

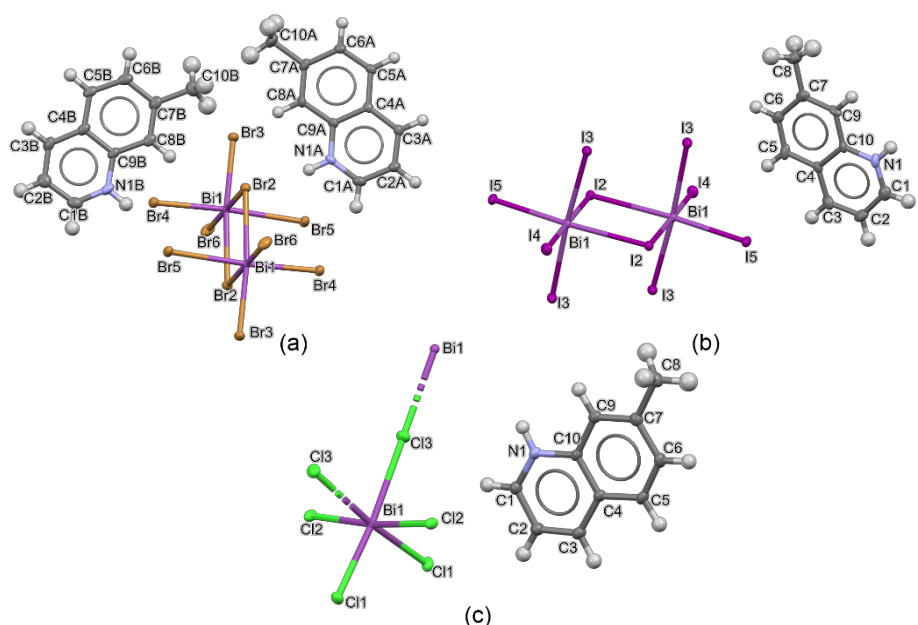

Figure 1. Structure building units a) 7-MeqBiBr$_3$ with [Bi$_2$Br$_{10}$] units; b) 7-MeqBiI$_3$ with [Bi$_2$I$_{10}$]; c) 7-MeqBiCl$_3$ with [BiCl$_5$]$_n$ chains.

The three crystal structures contain a common cation: 7-methylquinolinium (7-Meq). The crystal data and refinement details are presented in SI file (Table S1) alongside the Bi-X bond lengths (Table S2). The intermolecular interactions involving the 7-methylquionoline cations are summarized in Table S3-S6. Increasing the temperature during synthesis from 100 °C to 140 °C resulted in a second iodine structure (7-MeqBiI$_3$-I) with one dimensional [BiI$_4$]$_n$ chain and a triclinic crystal structure. In monoclinic 7-MeqBiI$_3$ we observe 0D inorganic [Bi$_2$I$_{10}$] parts and organic cations, which are disordered over two positions with occupancies 0.62:0.38.

The 7-MeqBiBr$_3$ crystallizes in the triclinic crystal system with the P-1 space group. Within this structure, we can identify discrete Bi$_2$Br$_{10}$ motifs, with each asymmetric unit containing half of this motif along with two organic cations. The organic components are held together through π⋯π interactions, forming stacks as detailed in Table S6. When examining the structure, we observe the formation of a brick wall motif created from both types of components, particularly when viewed along the a-axis (refer to Figure S1). In this crystal structure, the inorganic parts are connected to the organic entities through a series of intermolecular interactions, with the strongest interaction being of the N-H···Br type, resulting in the formation of a $D_1^1(2)$ motif. It's important to note that there are no direct hydrogen bonds observed between the 7-methylquinolinium cations, and there are also no halogen bonds present within this particular crystal structure (Figure S2). The structure of 7-MeqBiCl$_3$ belongs to the monoclinic, centrosymmetric C2/c space group. Within this structure, there are infinite zig-zag like chains of [BiCl$_5$]$_n$ along [001] direction. These adjacent chains are interconnected through the presence of organic cations. Notably, the strongest interaction observed is N-H···Cl. It is important to highlight that the Bi-I interactions within one octahedron are not equal in strengths (distances vary from 2.561Å to 2.886 Å). Furthermore, the Bi cation is positioned on a 2-fold axis, while the two octahedral I ions that bridge it are situated on a mirror plane. Importantly, there are no π···π interactions between the organic cations, with the shortest Cg···Cg distance being approximately 4.53Å.

7-MeqBiI$_3$-I crystallizes in triclinic P-1 space group. The inorganic part of the structure consists of one-dimensional [BiI$_6$]$_n$ chains that propagate along [100], formed by edge-sharing Bi$_2$I$_9$ moieties (Figure S3). This structural motif is not uncommon in iodobismuthates, as demonstrated in previous studies [56, 57]. The Bi-I bond distances within those chains vary from 2.910Å to 3.300Å. The shortest distance between the centroids of two phenyl rings of quinoline, related by the inversion centre (2-x, 2-y, -z) is 3.765Å. Furthermore, the inorganic chains are connected by a set of intermolecular interactions of C-H···I and N-H···I type. Fingerprint plots calculated from Hirshfeld surfaces revealed the subtle differences in the interactions between the organic ligands in those three crystal structures. In particular the diffused set of points that vary between the plots (Figure S4) come from the C···H and H···H contacts. Structures with 1D motif show more distinct halogen-hydrogen interactions, while discrete 0D units display more varied and diffuse patterns from multiple interaction types. The percentage contribution of each interaction can be seen in Figure S5.

The series of prepared samples at 100°C were further analyzed by measuring UV-Vis absorption in dimethylformamide solvent and diffuse reflectance spectroscopy (DRS) of solid samples which are fundamental techniques in semiconductor research, enabling the optimization of materials for optoelectronic devices, where precise control of optical properties is critical. Figure 2a display maximum absorption bands shifts from 484 nm in 7-MeqBiI$_3$ (red crystals) to 376 nm in 7-MeqBiBr$_3$ (yellow crystals) and 320 nm in 7-MeqBiCl$_3$ (colorless crystals) respectively. These bands can be assigned to LMCT (halogen to bismuth) transition, the change in energy of this transitions clearly follows the changes in electronegativity of the ligand. The same trend is visible in diffuse reflectance spectra of solid samples.

A more detailed analysis of diffuse reflectance spectra has been performed based on Tauc method. Firstly, diffuse reflectance spectra were converted to the Kubelka-Munk function, defined as follows (1):

$$F(R) = \frac{(1-R)^2}{2R} \qquad (1)$$

where *R* is the reflectance. For powder samples dispersed in scattering media it is commonly assumed that F(R) is proportional to the absorption coefficient α. Then the Tauc function has been applied to fit the linear fragment of the spectrum (2):

$$(\alpha \cdot hv)^{\frac{1}{n}} = A(hv - E_g) \qquad (2)$$

where *A* is the proportionality constant (independent of the photon energy), *h* is the Planck constant, *v* is the photon frequency, $E_g$ is the band gap, α is the absorption coefficient, and *n* is the factor describing the nature of the band gap. In the case of amorphous (or almost amorphous) material it is assumed that *n* = 1 seems to be the most reasonable choice [58]. The same approximation is commonly used for molecular crystalline materials as well as for ionic crystal with only a weak covalent interaction between ionic species [59, 60]. Even though DFT models (*vide infra*) predict indirect band gap the n = 1 case provides the best fit for the spectra. Therefore, the final equation that allows the determination of the band gap is derived as follows (3):

$$F(R) \cdot hv = A(hv - E_g) \qquad (3)$$

The obtained band gaps were $E_{g(7-\text{MeqBiI}_3)} = 2.09$ eV, $E_{g(7-\text{MeqBiBr}_3)} = 2.77$ eV, and $E_{g(7-\text{MeqBiCl}_3)} = 3.19$ eV, respectively.

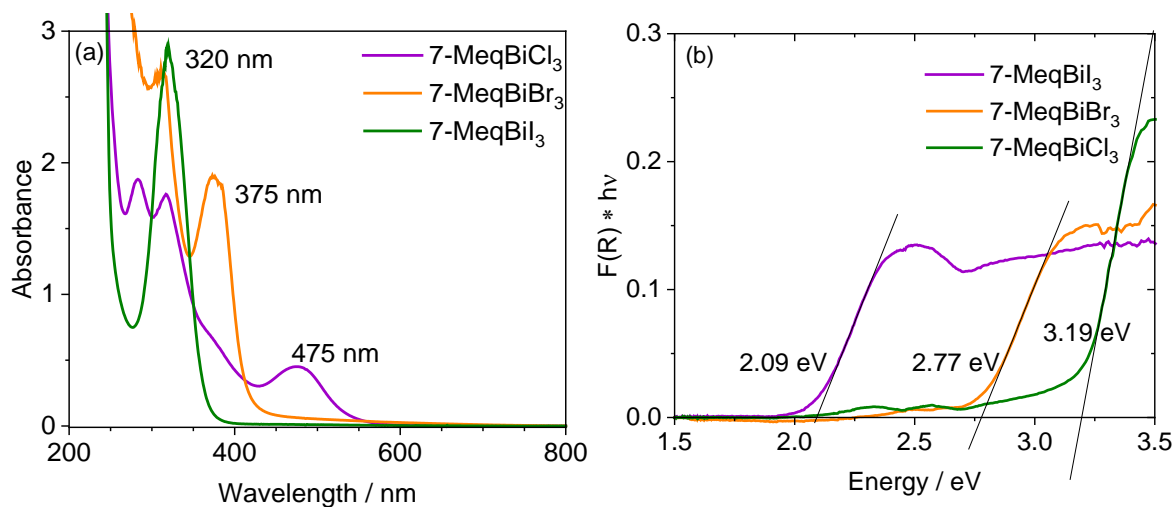

Figure 2. a) UV-Vis absorption spectra of complexes in dimethylformamide solutions and b) Tauc plot on the basis of the Kubelka-Munk function.

The band structure diagrams, along with the DOS and PDOS distributions calculated using DFT, are presented in Figure 3.

The upper part of the valence band of all three compounds is built from 6p orbitals of iodide ligands, with a small admixture of some carbon centred orbitals, especially if it is lower energy range. The conduction band is split into two sections – lower, composed mostly of carbon-based orbitals with a small contribution of 3p states of nitrogen and the higher energy zone, with a significant contribution of bismuth-centred orbitals. These results are quite similar to band structures of other organic halobismutates [21, 31, 61] as well as mixed-ligand bismuth complexes [28]. It is in striking contrast to leading halide perovskites (and other perovskite semiconductors), in which the lower part of the conduction band is mainly composed of metal-centred orbitals. The peculiar band structure: localization of valence bands mainly on anionic sublattice and conduction band on cationic sublattice results in weak of bands and in consequence results in low charge carrier mobilities. Therefore most of the halobismutate salts, with some notable exceptions [62, 63], despite their satisfactory stability and optical properties are not suitable for photovoltaic applications, or show only very low efficiency [64]. This, however, does not disqualify them for other applications, where low mobility (and resulting lower conductivity) may be considered beneficial. These include memristive properties, where bismuth iodide derived materials have already demonstrated their good and diverse performance [29-31, 46, 65].

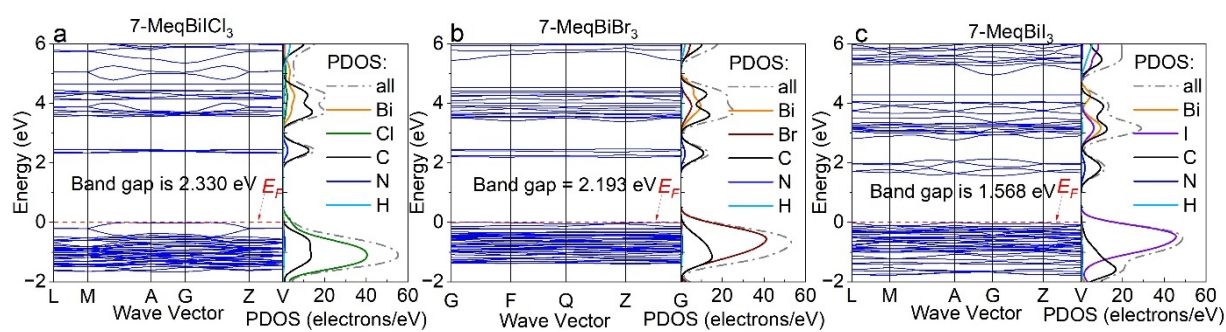

Figure 3. Left side: DFT-calculated band structures of (a) 7-MeqBiCl$_3$, (b) 7-MeqBiBr$_3$, (c) 7-MeqBiI$_3$ collated with the right side: partial density of states (PDOS).

Unfortunately, 7-MeqBiCl$_3$ and 7-MeqBiBr$_3$ were not soluble in any of the available solvents and not volatile, which prevented preparation of homogenous layers and fabrication of devices. The only memristor devices based on compounds from the studied family comprised of thin layers of 7-MeqBiI$_3$ sandwiched between ITO glass as the counter electrode and evaporated copper top electrode serving as the working electrode. Spectroscopic ellipsometry was used to determine the thickness of the materials. Measurements were taken at incident angles ranging from 50° to 70° in 5° increments, as illustrated by Ψ and Δ in Figure S6. The adoption of a biaxial anisotropic model with Tauc-Lorentz layers was applied for fitting the experimental data.

### 3.2 Memristive properties of the prepared devices

The quality of the layers as well as the homogeneity of the material has been investigated using SEM imaging and energy-dispersive X-ray spectroscopy (EDX) (Figure 4). SEM images of layers prepared from concentrated solutions are of sufficient quality. Despite some visible changes in contrast, distribution of all key elements is fully homogeneous and no irregularities in stoichiometry were found. The cross-sectional SEM image revealed a thin layer with a thickness between 100-130 nm (Figure 4d). Notably, solutions with concentrations lower than that of 150 mg of 7-MeqBiI$_3$ in 1 mL dimethylformamide were too diluted to adequately cover the roughness of the ITO surface, potentially causing short circuits during electrical measurements.

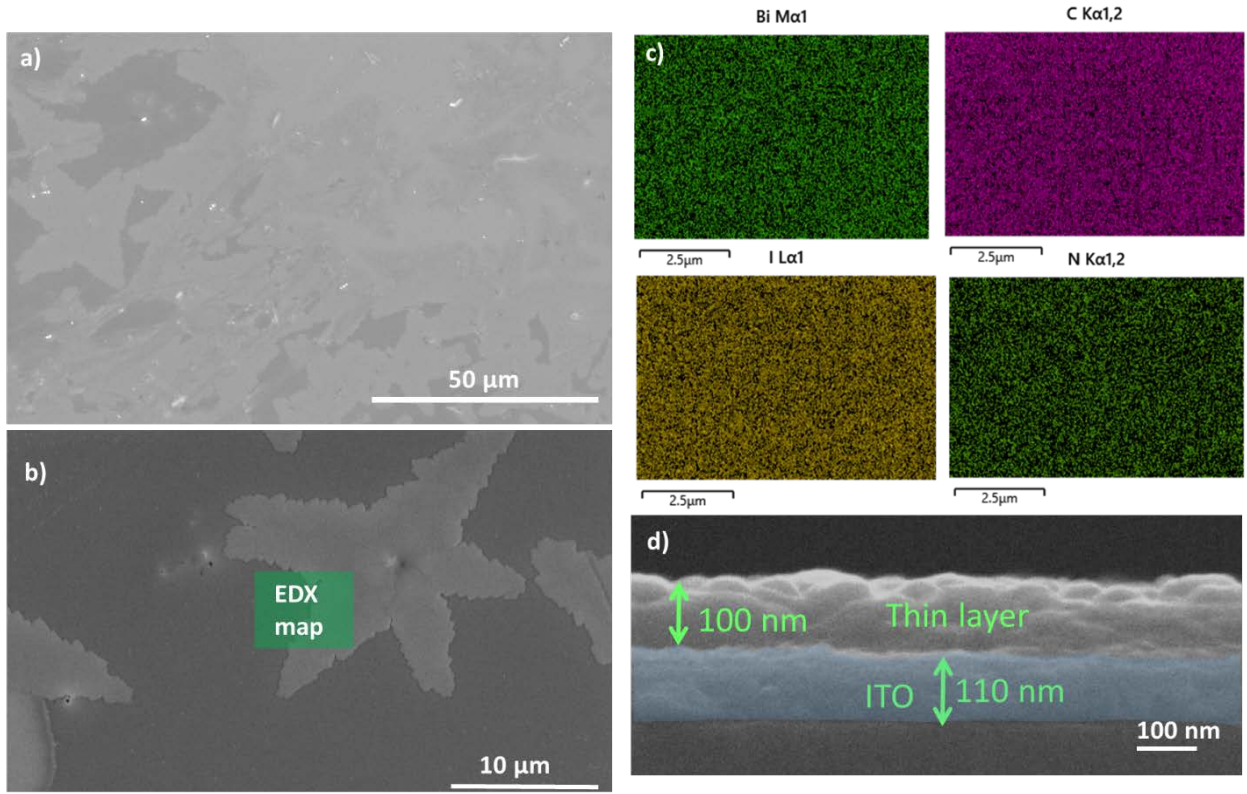

Figure 4. SEM analysis of 7-MeqBiI$_3$ on ITO/glass: a) and b) spin coated thin layer with different resolution c) elemental mapping of Bi, C, I, and N using EDX spectroscopy, d) cross-section of thin film onto 110 nm of ITO on glass.

In memristive devices, made of 7-MeqBiI$_3$ sandwiched between ITO and Cu electrodes, a voltage sweep test was applied to analyze the current-voltage (I-V) characteristics at scan rates of 25, 250, and 500 mV/s across a range of voltages as ±0.5, ±1 and ±2 V, respectively (Figure 5). These tests are critical for observing the resistive switching behavior, particularly the response of the device under test to different voltage ranges and the time-dependence of the switching process. A hysteresis loop appears in the ±1 V scan range, indicating threshold-driven switching likely caused by ion migration or charge trapping, while no loop is observed at smaller voltage ranges s (±0.5V), suggesting that the device activates only beyond a specific threshold. When forward bias represents low resistance and moving to high resistance in the first quadrant usually indicates the resetting of a memristive device from a low resistivity state (LRS) to a high resistivity state (HRS), marking a shift from ON to OFF. The loop is asymmetric, with a larger clockwise loop in the first quadrant (+V, +I) and a more linear response with lower resistance in the third quadrant (-V, -I), indicating at least partially rectifying character of the junction. This may suggest formation of Schottky barrier at 7-MeqBiI$_3$ – copper interface. The performance of the device (i.e. the shape of hysteresis loop) does not change significantly within studied scan rate values (25-500 mV/s). This behavior, which is not in line with classical description of the memristor may suggest that diffusion processes, which are relatively slow compared to applied scan rates, do not play any significant role in resistive switching in ITO|7-MeqBiI$_3$|Cu devices, however, may contribute to slow changes in ON and OFF currents

affecting the state retention profiles (Figure 5d). Alternatively, we can postulate the interfacial mechanism, related to modulation of the Schottky barrier height, at the main mechanism of resistive switching in this case. Asymmetry of hysteresis loop (the forward-to-backward current ratio of ca. 2), as well as relatively low ON:OFF ratio (*vide infra*) also corroborates this hypothesis. This memristive device likely operates through interface-driven effects at the ITO and Cu electrodes, with a voltage and time-dependent threshold switching behavior, making it suitable for neuromorphic computing applications. Namely, copper electrode forms a Schottky junction with 7-methylquinolinium iodobismuthate. The metal-induced gap states are then formed at the interface region [66, 67]. They can be populated and depopulated at appropriate threshold potentials and thus modulate the Schottky barrier height. Mixed conductivity character, with minor contribution of ionic conductivity (cf. Figure 6, *vide infra*) corroborates this hypothesis.

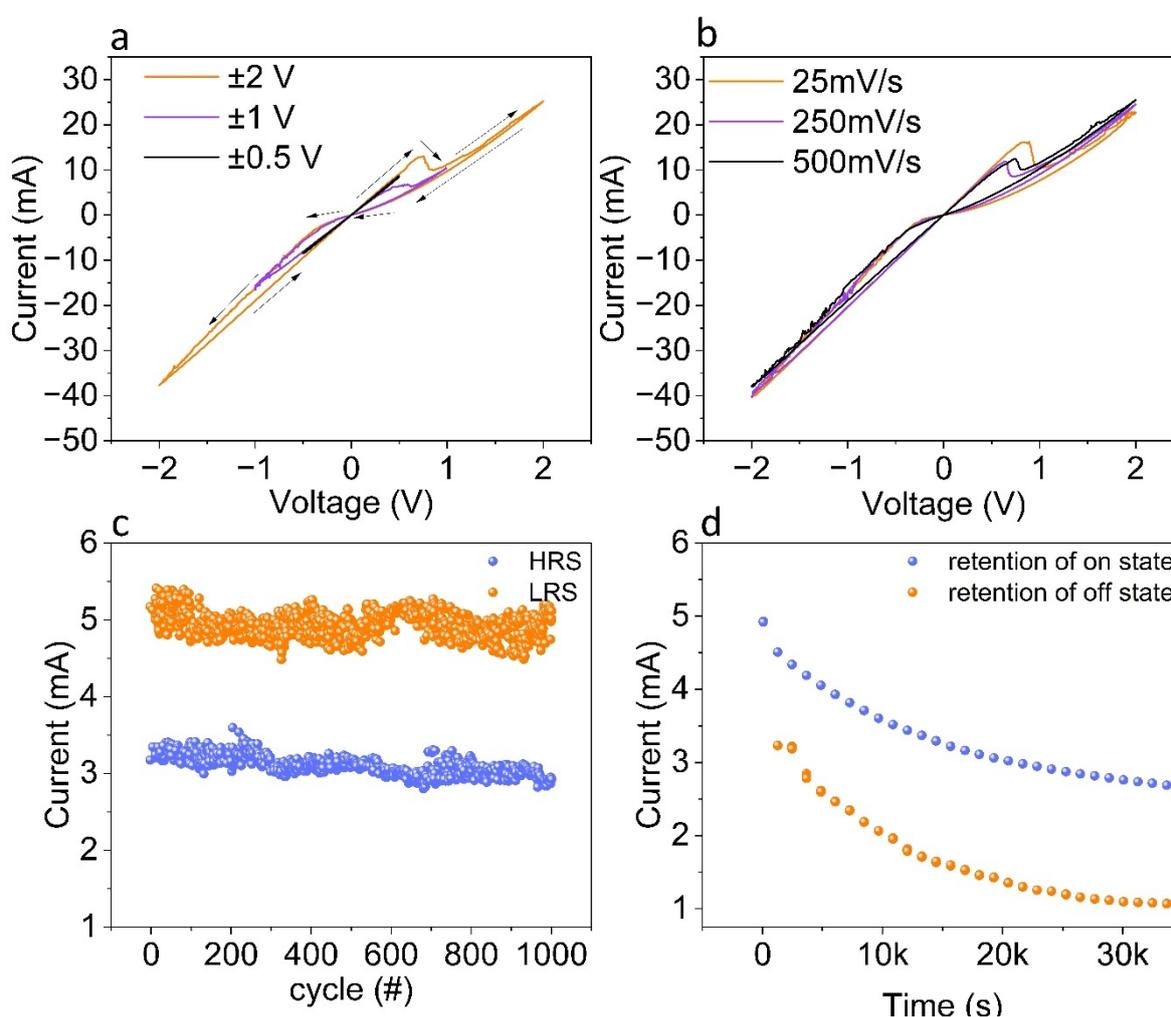

Figure 5. I-V scan of 7-MeqBiI$_3$ at a) different ranges, b) different scan rate c) ON-OFF measurement with set pulse (amplitude -1 V and width 100ms) and reset pulse (+1V and width 100ms) and reading point at 0.3 V, d) retention of ON and OFF states (i.e. LRS and HRS, respectively) over more than 11 h.

For the ON-OFF test, a sequence of pulses was applied to the memristive device, starting with a +2V (reset) and a pulse width of 0.1 seconds, followed by reading the state at 0.3V (Figure 5c, Figure S7). This was then followed by a -2V pulse (set) of the same duration (0.1 seconds), after which the state of the device was again read at 0.3V. This pulsing sequence was repeated for a total of 1000 cycles. The results indicated that the ratio of the ON state (low resistance state, LRS) to the OFF state (high resistance state, HRS) was less than 2, demonstrating the device's ability to switch between these two states effectively and reproducibly. A comprehensive performance comparison of resistive switching (RS) devices based on various organic and inorganic lead-based and lead-free materials is presented in Table 1.

**Table 1.** Comparison of performance of Cu/7-MetqBiI$_3$/ITO memristor with other recently reported memristors.

| Device | $V_{set}$ | Retention (s) | On/Off ratio | Thickness | Endurance (#) | Ref. |
|---|---|---|---|---|---|---|
| Au/PMMA/Cs$_3$Bi$_2$I$_9$/ PEDOT:PSS/ITO | +0.73 | $10^5$ | $4.3 \times 10^3$ | 165 nm | $10^4$ | [114] |
| Ag/(CH$_3$NH$_3$)$_3$Bi$_2$I$_9$/ITO | 3 | $1.5 \times 10^3$ | >10 | 900 nm | - | [115] |
| Ag/PCBM/Cs$_3$Bi$_2$I$_9$/PEDOT:PSS/ITO | - | $4 \times 10^4$ | $> 10^5$ | 1 $\mu$m | $10^4$ | [116] |
| Au/α-CsPbI$_3$ QDs/PEDOT: PSS/ITO | 0.5 | $4 \times 10^4$ | $8 \times 10^4$ | 250 nm | 800 | [117] |
| Ag/CsSnCl$_3$/ITO | 0.95 | $>10^4$ | $10^2$ | - | $10^5$ | [118] |
| Al/Cs$_2$AgBiBr$_6$/ITO | -0.5 | $10^4$ | $10^3$ | - | $> 2\times 10^3$ | [119] |
| Ag/Bi/MAPbI$_3$/ITO | 1 | $10^4$ | $10^2$ | - | 800 | [120] |
| Ag/Bi$_2$S$_3$/ITO | −0.58 | $>15 \times 10^3$ | - | 90 µm | $6\times 10^3$ | [121] |
| Ag/Sb$_2$S$_3$/Pt | 0.3-0.4 | $> 10^4$ | 200 | 191 nm | $10^3$ | [122] |
| Au/Cs$_2$AgBiBr$_6$/ITO | +1.53 | $10^5$ | 10 | 180 nm | $10^3$ | [123] |
| Au/MASnBr$_3$/ITO | ~0.65 | $10^4$ | $10^2$ -$10^3$ | 2 µm | $10^4$ | [124] |
| Ag/PMMA/Cs$_3$Cu$_2$I$_5$/ITO | +1 | $>10^4$ | $10^2$ | 500 nm | 100 | [125] |
| Au/PMMA/CsSnI$_3$/ITO | +1 | $10^4$ | $10^3$ | 300 nm | $1.5 \times 10^2$ | [126] |
| Au/(CH$_3$NH$_3$)$_3$Bi$_2$I$_9$/ITO | +1.6 | $10^4$ | 100 | 250nm | $3 \times 10^2$ | [127] |
| Ag/PMMA/Cs$_2$AgBiBr$_6$/ITO | +0.5 | $10^3$ | 10 | 240nm | 100 | [128] |
| Au/Rb$_3$Bi$_2$I$_9$/Pt/Ti/SiO$_2$/Si | +0.5 | $10^3$ | $2.9 \times 10^7$ | 160nm | 200 | [129] |
| Cu/[Ni(Me$_4$dtaa)]/ITO | 1.5 | >500 | <2 | 150 nm | >500 | [70] |
| Cu/TPCPD/ITO | 0.8 | $>4\times 10^4$ | 1.3 | 120 nm | $>3.5\times 10^5$ | [130] |
| Al/𝛿-FAPbI$_3$/ITO | 2.6 | $10^3$ | $>10^5$ | - | 250 | [131] |
| Cu/4-CNpyBiI$_3$/ITO | -2 | $> 3\times 10^4$ | 13 | 200 nm | 100 | [132] |
| Cu/7-MeBiI$_3$/ITO | -2 | $> 3\times 10^4$ | <2 | 100 nm | 1000 | This work |

Abbriviations: PEDOT: poly(3,4-ethylenedioxythiophene); PSS: poly(styrenesulfonate); PMMA: Polymethyl methacrylate; MA: methylammonium; TPCPD: tetraphenylcyclopentadienone; [Ni(Me4dtaa)]: nickel-7,16-dihydro-6,8,15,17-tetramethyldibenzo[b,i]−1,4,8,11-tetraazacyclotetradeca-4,6,11,13-tetraene; FA: formamidinium cation; 4-CNpy: 4-cyanopyridinium.

Two types of devices with different surface area of top electrodes as 1 mm$^2$ and 9 mm$^2$ were prepared and the measurements of IV plots for ten devices were performed. ON and OFF states for 100 cycles in each device were calculated. Results are presented in Figure S8 and Table S7, and prove the increased current in ON and OFF states both, by increasing the size of the electrode which is in agreements with our primary inference from asymmetric IV curve. The lack of a sharp transition leans more toward interface-dominated switching rather than a pure filamentary effect. Log-log plots of current (I) versus voltage (V) in resistive switching devices reveal essential information about charge transport, facilitating a detailed analysis of conduction mechanisms (Figure S9). Different transport phenomena, including Schottky emission, space-charge-limited conduction (SCLC), and trap-assisted tunneling (TAT), produce distinct slope behaviors. By examining these variations, researchers can identify conduction regimes and gain deeper insights into the fundamental processes driving resistive switching. SET process follows trap-controlled SCLC conduction, while the RESET process is more abrupt, likely due to interface barrier modulation, as it changes with surface area of the electrodes. Observed electrical characteristics: the absence of a significant dependence of switching characteristics on scan rate (Figure 5 b), low ON/OFF ratio and small but distinct rectification factor of ca. 2 indicate that the primary switching mechanism involves charge trapping at the metal semiconductor interface, which in turn modulates Schottky barrier height. Positive polarization of the working electrode (copper top contacts) results in depopulation of metal-induced gap states (MIGS) within the band gap (Figure 6a) [66, 67]. This results in an increase of Schottky barrier height, further increased migration of iodide ions (or iodine vacancies) can contribute to this process as well. In the high resistance state the Schottky barrier is high due to the downward band bending at the interface (Figure 6a). Negative polarization of the top electrode results in partial occupation of metal-induced gap states, therefore the band edges assume higher energy. In the case of *p*-type semiconductor it results in decrease if the Schottky barrier height (Figure 6b). The process is fully reversible, and the rates of population and depopulation of trap states are high (much higher than any ionic movements in solid complex). Due to polycrystystalline nature of the film, the rate of trapping and detrapping of electrons at MIGS, although relatively high, is not uniform at the whole surface of the electrode. The same concerns the energy of MIGS – due to polycrystallinity we should expect Gaussian distribution of their energies, like in the case of surface states at other interfaces [68, 69], therefore hysteresis is observed. Furthermore, trapping and detrapping of electrons results in partial capacitive character of the junction [70]. Therefore, the equivalent circuit of the device is not straightforward (Figure 6c). It encompasses bulk resistivity of the material ($R_1$), ON and OFF resistances of the Schottky junction at two states (shown as ideal memristor with two resistances, namely $R_{ON}$ and $R_{OFF}$), an ideal Schottky barrier diode D and a capacitor C. This electrical complexity of a single junction has further consequences for computational performance of studied material (*vide infra*).

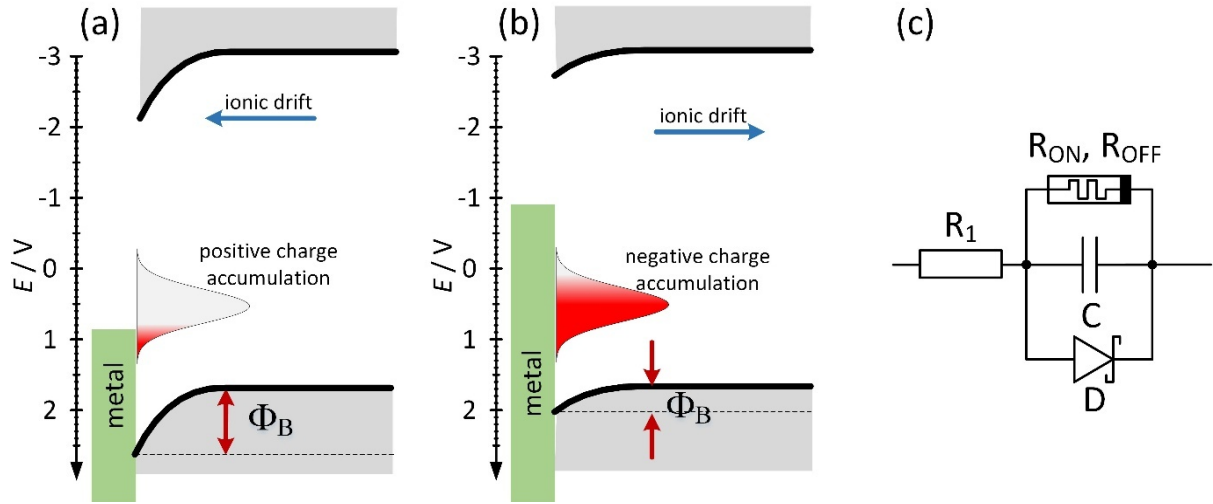

Figure 6. A tentative mechanism of resistive switching in Cu|7MeqBiI$_3$|ITO device: electronic structure diagram for the HRS (a) and LRS (b). A tentative equivalent circuit for the junction (c).

The behavior of the memristive device across various temperature regions (from -30 to 120 °C) at +1 V and -1 V reveals complex conduction mechanisms (Figure 7). In the first region (approximately −10 °C to 25 °C), the device exhibits a positive slope, related to some thermal scattering of charge carriers. Conductivity abruptly increases (for both HRS and LRS states) within a narrow temperature range (25 to 50 °C). Linearized Arrhenius equation has been used for determination of activation energy of the process responsible for an increase in conductivity (4):

$$\ln \sigma = A - \frac{E_a}{k_B}\frac{1}{T} \qquad (4)$$

where $E_a$ is the activation energy and $k_B$ is the Boltzmann constant. Calculated activation energies are 0.26 (25.1) and 0.31 eV (29.9 kJ/mol) for +1 and −1 V bias respectively. These voltage values correspond to the device fully switched to HRS and LRS states, respectively (cf. Figure 5a). Very similar activation energies for HRS and LRS indicate only minor (if any) contribution of ionic conductivity to resistive switching processes. This is in line with a preliminary conclusion (on the basis of voltammetry analysis) that interfacial switching is a dominating mechanism for 7-MeqBiI$_3$-based devices. These values are somehow lower than values obtained for lead halide perovskites (0.33-0.58 eV) [71, 72], but higher than in the case of other iodobismuthates [21, 31]. This process can be attributed to vacancy-induced iodide diffusion, which is a common process in halide ionic semiconductors [73]. It should be also noted that electron scattering increases with increasing temperature both below and above the threshold temperature range with negative slope (which corresponds to activation of ionic transport). In indicates stronger scattering at higher temperature, also indicating mixed electronic and ionic conductivity in this material. Such a feature may be beneficial for neuromorphic applications, especially due to non-linear character and possibility of a complex frequency response.

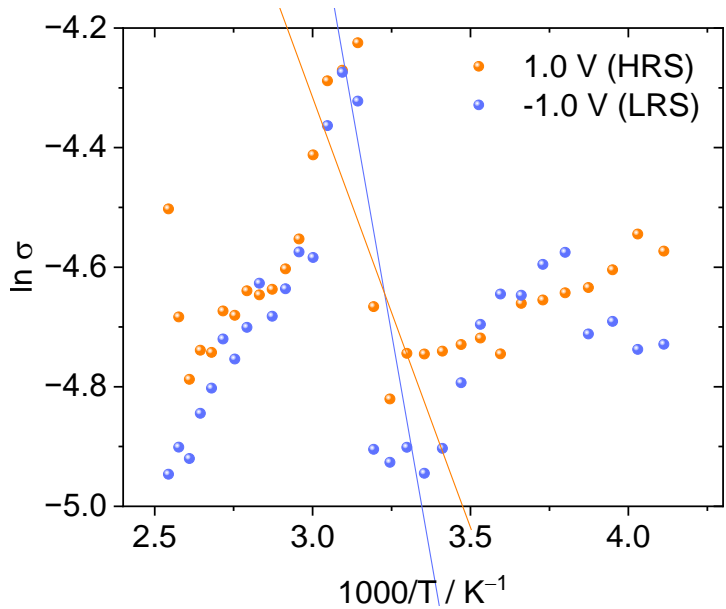

Figure 7. Arrhenius plot calculated from conductivity measurement at −1V and +1 V within the −30 - +120 °C temperature window.

The potentiation-depression plot for the memristive device was measured using 100 pulses of a range of amplitude (-2.6 to 0.4 with 0.2 V interval) with widths of 1 and 10 ms. The device state was read at 0.3V after each pulse. Following this, the device was sequentially depressed back to the off state with 100 pulses of +2V to +0.4V amplitude. This test is crucial for evaluating the device's ability to transition between high and low resistance states, which is essential for applications in memory storage and neuromorphic computing. The ON state for the smaller pulse width showed lower conductance compared to the 10 ms pulses, likely because shorter pulses provided less time for the device to transition fully to the low resistance state (LRS). Shorter pulses may limit excessive ion migration or defect state formation, allowing the device to fully reset to the high resistance state (HRS). Longer pulse widths may encourage deeper ion migration or incomplete recovery, preventing full switching back to the HRS as it is shown in Figure 8. Thus, longer pulses result in more reliable transitions between ON and OFF states.

The STDP was implemented by applying pre- ($V_{pre}$) and post-synaptic pulses ($V_{post}$) at varying time intervals, with a sequentially positive and negative time difference ($\Delta t$). The process for initiating and resetting the device is according to the reported procedure. Figure 8c illustrates the changes in synaptic weights. The pulse time difference ($\Delta t$) significantly impacts the resulting voltage shape ($V_{pre}$—$V_{post}$), a key factor in determining final synaptic weights. When $\Delta t$ approaches zero, the device conductance reaches its maximum ($G_{max}$) due to the increased voltage from spike superposition. Conversely, a time difference exceeding the duration of the pulse, the conductance of the device is preserved. In this framework, the synaptic weight changes ($\Delta W$), defined as (5):

$$\Delta W = \frac{G - G_0}{G_0} \tag{5}$$

displays an asymmetric Hebbian pattern, peaking around $|\Delta t| \sim 0$.

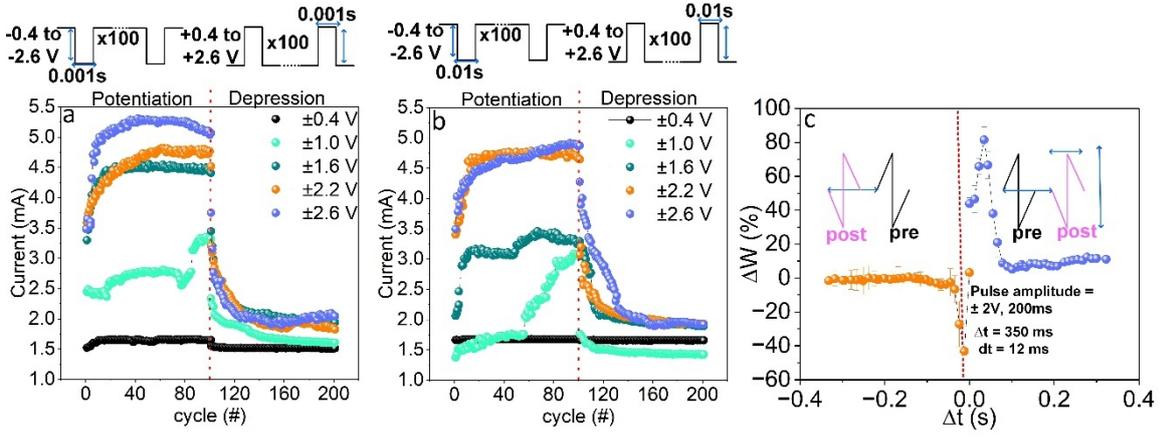

Figure 8. Potentiation and depression with different pulse amplitude and widths a) 1 ms and b) 10 ms, c) spike timing-dependent plasticity implemented with bipolar right-angled triangles with 200ms width and ±2 V amplitude.

### 3.3 Preparation of physical reservoir, characterization and application

One of the main strengths of reservoir computing is its ability to transform low-dimensional input signals into higher-dimensional outputs with different frequencies and waveforms [74, 75]. This can be demonstrated in a waveform generation task, where a sine wave was applied to an input of a physical reservoir, whereas 15 output signals were recorded and subsequently used for reconstruction of various target waveforms, including cosine, sawtooth, square, and triangle waves as various linear combinations (with coefficients/weights obtained via training procedure) of 15 output signals (6):

$$u_{predicted}(t) = \sum_{i=1}^{15} w_i\, u_{output,i} \tag{6}$$

Results are shown in Figure 9. The evaluation of weights for each target signal was achieved via minimalization of the normalized mean squared error (NMSE), as NMSE seems to avoid bias towards models that overpredict or underpredict the target waveform and emphasizes the scatter in the entire data set [76]. NMSE is defined as follows (7):

$$NMSE = \frac{\sum_{t=1}^{N}\left(y_{target}(t) - y_{predicted}(t)\right)^2}{\sum_{t=1}^{N}\left(y_{target}(t)\right)^2} \tag{7}$$

where $u_{target}(t)$ and $u_{predicted}(t)$ are values of obtained and desired signal amplitudes within the studied waveform.

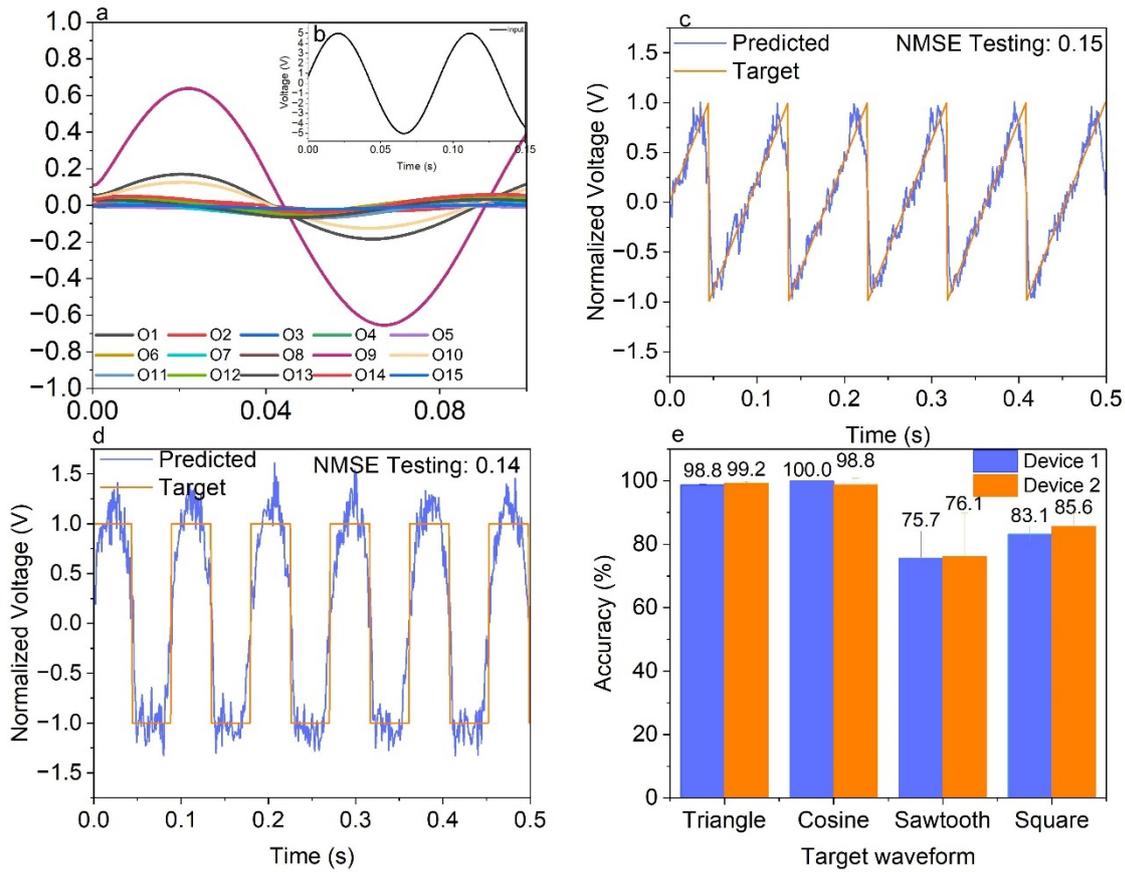

Figure 9. a) sine waves of 15 outputs, b) inset represents an input part. Mapping sine wave to c) sawtooth and d) square waveforms. e) Represent the accuracy for two different devices for four types of waves: square, triangle, cosine, and sawtooth.

The nonlinearity of the device can be further demonstrated by Fourier analysis of output signal recorded for low frequency sine wave inputs. The power spectral density (PSD) shows how signal power is distributed over frequencies and is derived by squaring the magnitudes of the fast Fourier transform (FFT) and normalizing. While the FFT gives frequency components and amplitudes, PSD quantifies power at each frequency. A log-log plot of PSD is often used to analyze noise characteristics across different frequency scales. Higher harmonics are the most straightforward way to visualize nonlinearity, which plays a crucial role in reservoir computing because it generates additional features contributing to increased dimensionality of transformed signals (data). In physical reservoir computing, the input signal is processed through a nonlinear physical system, called reservoir (possibly with rich internal dynamics), which maps the input into a higher-dimensional space [38, 77]. The presence of harmonic frequencies in outputs of the reservoir upon sine wave stimulation indicates that the system is introducing nonlinearities into the signal, enriching the representation of the input data. This nonlinear transformation allows the reservoir to capture intricate patterns, relationships, and dynamics that a purely linear system might miss. By leveraging harmonics, the reservoir can better separate, classify, or map input signals to target outputs, improving its ability to solve tasks such as waveform generation, speech recognition, or time-series prediction. In physical reservoirs, the generation of harmonics is often a sign of the system's rich dynamics and ability to process complex information through both linear and nonlinear interactions.

Upon application of DC voltage, the 7-MeqBiI$_3$ device (with one input and one output) generate noise the spectrum of which depends on applied voltage (Figure 10a and Figure S10). The noise spectra follow the power law (8):

$$PSD \sim \frac{1}{f^\beta} \qquad (8)$$

where $f$ is frequency and $β$ is the characteristic exponent. For white (thermal) noise $β = 0$, whereas pink noise (characteristic e.g. for filamentary processes) is described with $β = 1$. At zero bias the Cu/7-MetqBiI$_3$/ITO device generates white noise, but increased bias (either positive or negative) results in increased $β$ values up to ca. 1.5 (Figure 10). This indicates electrical instability of the junction which is demonstrated as rich chaotic dynamics of current across the junction. It may be associated with disorder in polycrystalline layer of active compound as well as chaotic movements of heavy ions within the lattice and resulting spontaneous fluctuations of the Schottky barrier. This behaviour may be also beneficial for reservoir computing applications, because it will provide a functional equivalent of masking function, which enriches the input data [78].

With these primary results and knowledge, we tried to evaluate the response of the physical reservoir device with one input-fifteen output to the applied voltages with very low frequency. The difference in harmonic ranges observed when applying 1V and 5V to the physical reservoir can be explained by the system's voltage-dependent nonlinear response (Figure 10b). At lower voltages, such as 1V, the reservoir behaves more linearly, resulting in a narrower harmonic range (up to 77 Hz). In contrast, at higher voltages, such as 5V, the system exhibits stronger nonlinear behavior, producing a much broader harmonic range (up to 319 Hz). This is due to the inherent nonlinear dynamics of the reservoir (resulting probably from mixed conductivity), where higher voltages drive the system into a more nonlinear regime, generating additional higher-order harmonics. The presence of nonlinear elements, such as memristive materials or non-Ohmic components, likely contributes to this behavior [79]. At higher voltages, these nonlinearities are more pronounced, leading to more complex oscillations and resonances within the reservoir. This behavior is consistent with passive higher harmonics generation in thin layer memristors resulting from multiple switching phenomena (formation of conductive pathways or local fluctuations of a Schottky barrier height) within the volume of the film [80]. The broader harmonic spectrum at 5V suggests that the system is capable of exciting higher-order modes and distributing energy across a wider frequency range, further demonstrating its nonlinear characteristics. This behavior is typical of systems with nonlinear dynamics and voltage-sensitive responses, where the harmonic content increases as the input voltage is raised. This nonlinear behaviour also qualifies studied devices for more advanced neuromorphic computing [81].

Another test of utility of a physical system for reservoir computing is analysis of the input-output dependence. Complex, nonlinear dependencies are a good symptom, indicating sufficiently rich internal dynamics which is beneficial for reservoir computing [2, 82, 83]. A simple implementation of this test is analysis of input-output two-dimensional plots, or analysis of fractal dimensions of input and output signals [84]. Higher fractal dimensionality is also a good indicator. Here we focus on the simplest test based on input-output correlations.

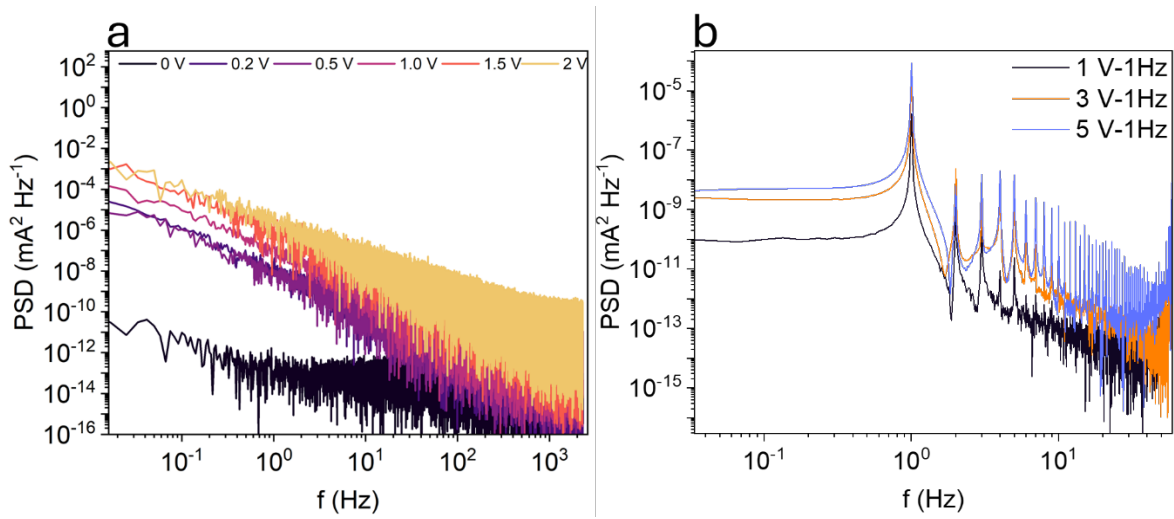

Figure 10. Power spectral density plots for a) ITO/7-MeqBiI$_3$/Cu device (one input and one output) in different DC voltages b) in different ac voltages with 1Hz for device with 16 electrodes (one input and 15 output) and b)

Plotting $V_{Output, j}$ ($V_{Oj}$) against $V_{Input}$ ($V_I$) in physical reservoir computing reveals how the system transforms signals, shedding light on its nonlinear dynamics and ability to map inputs into higher-dimensional spaces. This analysis represent the reservoir's capabilities in signal processing, memory retention, and nonlinearity, all crucial for tasks such as classification and prediction. The shape and complexity of the plot are key to assessing the reservoir's efficiency in distinguishing and transforming inputs for computation. An asymmetric or distorted ellipse which can indicate nonlinearity and capacitive behavior (which can be understood as a form of memory) [85] of that the reservoir can generate complex dynamics, have been observed (Figure 11), with three example of the $V_{O, j}$ ($V_{O, 3}$, $V_{O, 10}$ and $V_{O, 14}$) versus $V_I$. These plot shapes offer clear visual insights into the reservoir's potential to increase the dimensionality of the data space and its suitability for specific applications in physical reservoir computing. Here just visual inspection indicates that 1-dimensional inputs are mapped into two-dimensional phase space (Figure 11 a-b), however some output ports suggest one-dimensional character of the phase space (Figure 11 c). These data are consistent with at least partially capacitive character of junctions; however, some output ports are resistively coupled to the input. The irregular shape of the loop suggests more complex capacitive behavior, with a few capacitors (and parallel resistors) in series. This consistent with the structure of the device, as each electrode couple has two Schottky barriers (represented as an equivalent circuit comprising a diode, a resistor and a capacitor) [86, 87], as well as bulk capacitance of the material. When each $V_{O, j}$-$V_I$ plot differs across the outputs, it indicates that the reservoir is effectively expanding the input into a high-dimensional space with unique, nonlinear responses for each output. This diversity enhances the system's ability to differentiate inputs, supporting better signal separation and feature extraction. Such behavior is crucial for improving performance in tasks like classification and prediction.

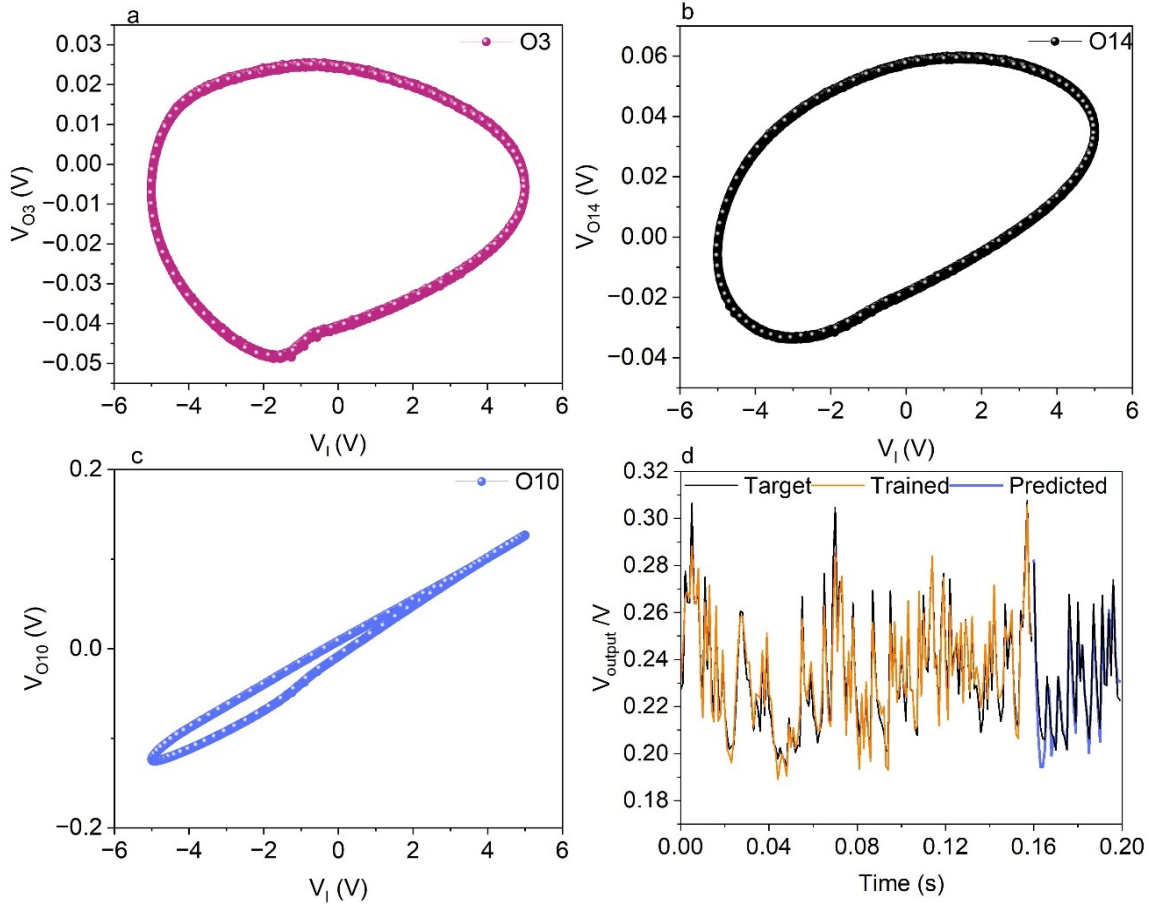

Figure 11. a-c) $V_{Output}$ ($V_{O,j}$) versus $V_{Input}$ ($V_I$) for output numbers O3, O14, and O10. d) the training and testing results for the NARMA2 task.

Another parameter used in the study of chaotic and dynamical systems is the correlation dimension ($\nu$) [84, 88]. It is used to probe dimensionality of the space occupied by a set of random points and is often referred to as a type of fractal dimension. For a finite time series of points described as Equation (9):

$$\{\vec{X}_i\}_{i=1} \equiv \{\vec{X}(t+i\tau)\}_{i=1} \qquad (9)$$

where $\tau$ is arbitrary but fixed time, increment the correlation integral is defined as (10):

$$C(r) = \lim_{N\to\infty} \frac{1}{N^2}\sum_{i,j=1}^{N} \Theta\left(r - \left|\overrightarrow{X(\imath)} - \overrightarrow{X(j)}\right|\right) \qquad (10)$$

where $\Theta(X)$ is a Heaviside step function. For small number $r$, the correlation integral behaves according to a power law (11):

$$C(r) \sim r^\nu \qquad (11)$$

where $\nu$ is interpreted as a fractal dimension or a trajectory in a phase space (here the input-output dependence) [88, 89]. Interestingly, the evaluated dimensionalities of input-output correlations (Figure 11) are consistent with noise exponents (Figure 10) from Eq. (8). This is not a coincidence. At very low voltages neither resistance nor capacitance play a role in the noise generation (white noise with $\beta = 0$ corresponds to zero-dimensional phase space with $\nu$

= 0), i.e. stochastic fluctuations dominate the dynamics of the system). With increasing voltage Ohmic processes dominate the transport phenomena in the material, which leads to pink noise and one-dimensional correlation exponent (β = 1 and ν = 1). Finally, at sufficiently high potentials charge trapping at the Schottky barrier takes place, which corresponds to memristive hysteresis, capacitive effects and "reddish" noise with β > 1 and higher dimensionality of the phase space (ν > 1). This observation implies that the performance of the reservoir based on the Cu/7-MetqBiI$_3$/ITO memristor depends on the amplitude of the input signal. This is consistent with previous studies showing very high input amplitude sensitivity of single-memristor reservoirs [90, 91].

The NARMA-2 (Nonlinear Auto Regressive Moving Average) task is a widely used benchmark in reservoir computing, designed to test a system's ability to handle temporal dependencies and nonlinear dynamics. The target output y(t+1) is computed by equation (12) [92, 93]:

$$y(t+1) = 0.4y(t) + 0.4y(t) \cdot y(t-1) + 0.6\, u^3(t) + 0.1 \qquad (12)$$

In this equation, the future output y(t+1) is a function of both past outputs and the nonlinear transformation of the current input u(t). This task is crucial because it evaluates a reservoir's memory retention and its capacity for performing complex, time-dependent computations, key requirements for tasks like time series forecasting and signal processing. Excelling at the NARMA2 task demonstrates that the reservoir can effectively balance short-term memory with nonlinear input-output relationships, enabling it to process dynamic data by integrating past and present information. First, an input sequence u(t) was generated using a uniform white noise signal with voltage fluctuation. A total epoch of 1 s was used for training and testing with a training ratio of 80 % of the data. Then Ridge regression as the final step for training the readout layer was applied [94]. Figure 11d illustrates the training and testing results for the NARMA2 task. Following the training process, the reconstructed output (depicted by the blue line) closely aligned with the target signal (represented by the gray line), yielding a calculated NMSE of 0.097. NMSE (Normalized Mean Squared Error) is a metric used to evaluate the performance of models, For the NARMA task, achieving an NMSE close to 0.1 or lower is often the goal, as it demonstrates effective modeling of the nonlinear relationships present in the task (13):

$$NMSE = \frac{\sum_{t=1}^{N}(y(t)-\hat{y}(t))^2}{\sum_{t=1}^{N}(y(t))^2} \qquad (13)$$

where $y(t)$ is the actual target output, $\hat{y}(t)$ is the predicted output from the model, and N the number of samples, respectively.

Memory capacity, implemented by applying random 5V rectangular pulses with different width and intervals (Figure S11), is typically evaluated by attempting to reconstruct the input signal from the reservoir state, with a delay of k time steps in the past. The total memory capacity (MC), is the cumulative ability of the reservoir to recall inputs over various time delays, i.e. (14):

$$MC = \sum_{k=1}^{K} MC_k \qquad (14)$$

where $MC_k$ is the memory capacity at delay k, and K is the maximum delay considered and defined as (15):

$$MC_k = \frac{cov^2(u(t-k), \hat{u}(t-k))}{var(u(t-k)) \cdot var(\hat{u}(t-k))} \tag{15}$$

The term $cov(u(t-k), \hat{u}(t-k))$ refers to the covariance between the input at time *t–k* and its predicted value. $var(u(t-k))$ and $var(\hat{u}(t-k))$ represent the variances of the actual and predicted inputs, respectively. The value of $MC_k$ measures how well the reservoir can reconstruct the input from *k* time steps in the past. The value of $MC_k$ measures how well the reservoir can reconstruct the input from k time steps in the past. The calculated total memory capacity is 35.23 for a maximum *k* of 100 delay steps meaning that the reservoir effectively retains and recalls information from approximately 35 time steps in the past (Figure 12a). In other words, the reservoir has meaningful memory of the input signal within this range, allowing it to reconstruct past inputs up to around 35 steps, with memory accuracy decreasing beyond this range.

This suggests that the reservoir has a significant ability to remember past information along with a fading memory feature [95, 96], which is essential for efficient implementation of tasks requiring short- to medium-term memory but may struggle with tasks requiring long-term dependencies, i.e. very long vectors of input data.

Thus, the device with only one input, which can take data in serialized form may have performance similar to artificial neural networks. The essential difference between data processing in reservoirs and neutral networks (this difference is crucial for processing of 2-dimensional data sets, e.g. images) is the structure of input layer and training procedures. Processing of 28 × 28 pixel image (like a character in MNIST [97] database) requires at least 786 inputs (one input for each pixel in black&white image, at least three times more for coloured images), which couple to the first hidden layer by 786 weights [98]. The optimal network for simple MNIST data sets classification consists of 4635 units, 98442 connections, and 2578 trainable weights [99]. Each of these weights needs to be independently trained. For reservoir, we can imagine a system with proper memory depth, so the data can be serialized and only one input may turn out to be sufficient. The output perceptron may minimally consist of only 10 output nodes (the number of nodes in equivalent to the number of categories), thus only 100-200 weights need to be trained in the minimal approach, depending on the number of output ports of the reservoir. This example shows the unprecedented simplicity of reservoir computing approach. This approach does not violate the "no free lunch theorem" [100-102], as the reduction of structural and training complexity of reservoirs versus artificial convolutional) neural networks is compensated by time required for computation: pixel-by-pixel processing of serialized data is much slower than parallel processing of all pixels at the same time [42], however may be favored from energetic point of view [103].

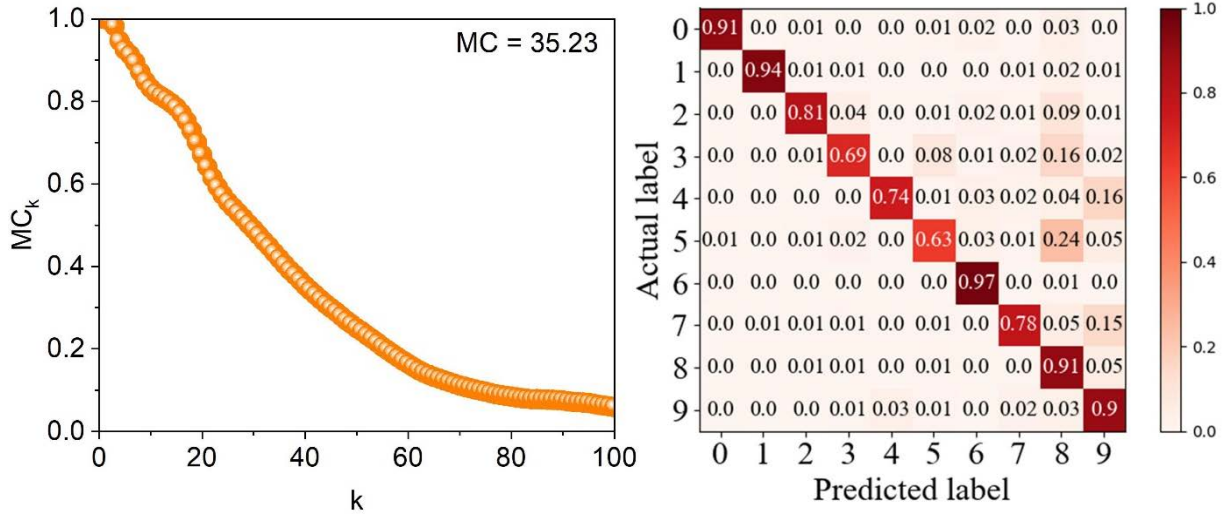

Figure 12. a) Memory capacity b) confusion matrix for digit classification.

The device designed with a single input and 15 outputs was employed for a classification task that involves recognizing handwritten digits from the MNIST dataset [97]. The input consists of 800 images for each of the 10 digits (0 to 9), resulting in a total of 8,000 images. Each digit image, which measures 28×28 pixels, was transformed into a 1D vector **u** of size 784 (fractional numbers considering shades of gray in original NMIST images). This 1D representation was fed into the device, with responses collected from the 15 outputs. The product of the input vector **u** by the trained weight matrix **W** (of 784 × 10 dimensionality) yield 10-element vector **z** containing the result of classification (16):

$$\mathbf{z} = \mathbf{u} \times \mathbf{W} \tag{16}$$

The individual elements of the output vector can be calculated as a Cauchy products according to (17):[104]

$$z_i = \sum_{k=1}^{784} u_k \cdot W_{k,i} \tag{17}$$

In this configuration, the perceptron functions as the readout layer, processing the 15 outputs from the device, each representing a transformed image, and mapping them to one of the 10-digit classes. This readout layer is comprised of 10 output nodes, with each node corresponding to a particular digit (0–9). Other words, the perceptron generates a vector of probabilities for each class in 10-shot classification. In order to improve the performance of the system the resulting output vector is subjected to final transformation by *softmax* function. It maps a vector **z** of *K* real numbers into a *K*-dimensional normalized vector, each component of which falls within the (0, 1) interval. Thus, the *softmax* function performs the mapping (18):

$$\sigma: \mathbb{R}^K \to (0,1)^K \tag{18}$$

for any *K* > 1 and is defined as (19):

$$\sigma(\mathbf{z})_i = \frac{e^{z_i}}{\sum_{j=1}^{K} e^{z_j}} \tag{19}$$

where **z** is the input vector. The *softmax* function amplifies any local maxima in the input vector, thus facilitates the one-hot classification of data.

The training entails a forward pass where the inputs are weighted and summed, followed by the application of the *softmax* function to produce output probabilities (Figure 12b). The model quantifies the loss using cross-entropy, which measures how well the predicted probabilities correspond to the actual one-hot encoded target vectors. The weights are then adjusted through backpropagation using optimization techniques like gradient descent to minimize the loss iteratively. Initially, 700 out of 800 handwritings were trained for 1,000 epochs, achieving an accuracy of 72% for this task. After several rounds of training, the number of epochs was extended to 5,000 and tested with the remaining 100 handwritings, leading to an accuracy improvement up to 82.26% (Figure 12). However, subsequent increases in the number of epochs resulted in diminishing improvements, suggesting that the model had reached a saturation point regarding classification performance. In comparison, the same perceptron, when trained on unprocessed vectors (i.e. working in linear approximation without reservoir) achieves maximum accuracy of 74 %. This clearly indicates that nonlinear transformations of image vectors results is some feature extraction and provides a significant improvement of classification tasks (ca. 20 % increase in classification accuracy) without any additional significant computational cost.

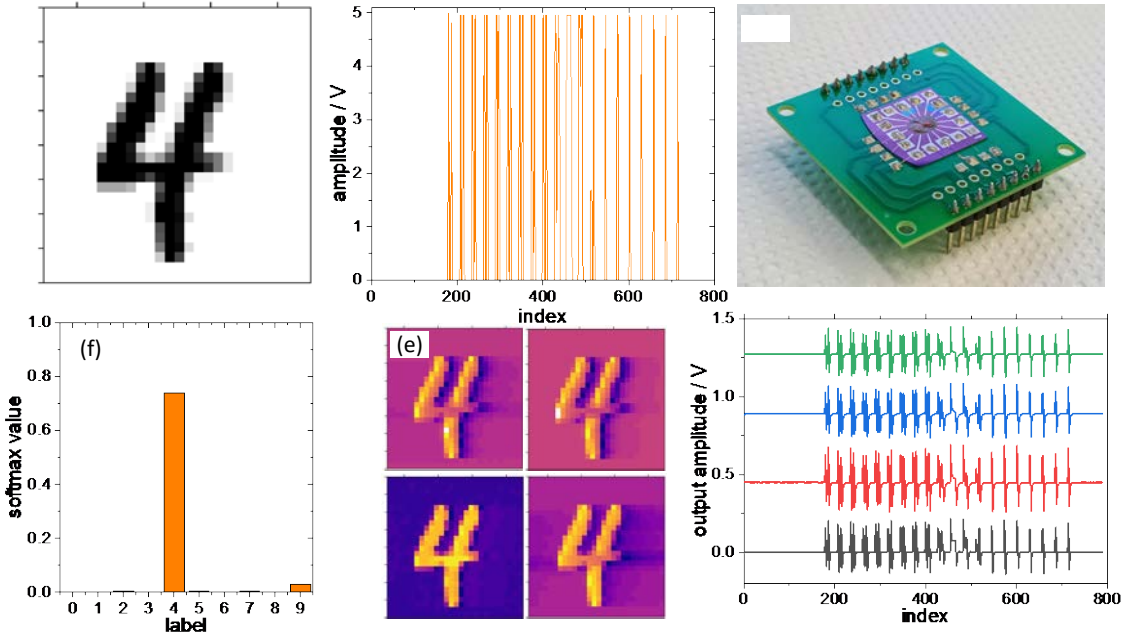

Figure 13. Schematic illustration of applied device in MNIST digit classifications: an example of input image (a) and its serialized version (784-element vector, b). A photo of a reservoir chip mounted on printed circuit board and wire-bonded (c). Examples of recorded outputs (d) reconstructed processed images (e) and resulting classification result (f).

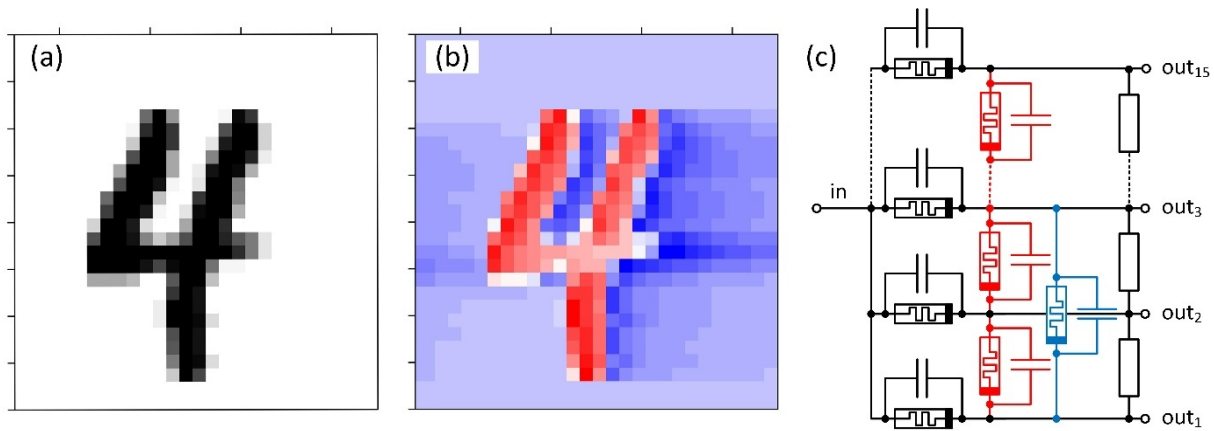

Figure 14. Signal intensity maps for one, randomly selected MNIST image: an input (a) an output with partial signal phase reversal (b) and a purely resistive output (c). Positive voltages are shown in red, zero is green and negative potentials are blue-colored.

Detailed analysis of signals recorded in reservoir reveals important details affecting their performance. Even from Figure 13 it can be seen that images (reconstructed from input and output vectors) undergo significant changes. The most striking change is the reversal of potential following the positive signal corresponding to the input (seen in the case of most of the outputs) (Figure 14a, b). We can clearly see that for some outputs a phase reversal is observed. This effect can be attributed to the capacitive components of the Schottky junction (cf. Figure 6c), or pseudocapacitance related to ion displacement within the material. Within a more complex circuit with several different resistance and capacitances (like in the studied case, where the active layer is polycrystalline and of non-uniform thickness), pulsed stimulation may lead to phase reversal. Charging and discharging currents have opposite values, therefore voltages across some of the resistors in the network also change their polarity. Such effect have been observed before for photoelectrodes made of non-homogeneous layers of nanoparticles [86, 105, 106]. Therefore, the device can be regarded as a complex (probably stochastic) network of memristive elements and capacitors. Such networks are known for complex electrical behavior [107, 108], nonlinearity and ability to transform electrical signal in a way facilitating their classification [84, 109]. This effect, which significantly enriches the data thus increasing the dimensionality of the data space and facilitating the classification task, originated form the complex capacitive structure of junctions. This is also similar to hyperpolarization of neuronal membrane following the action potential spike [110]. The second important feature is short term memory phenomenon. Memory capacity, evaluated to be ca. 35 time steps (Figure 12, equations 14-15) is clearly visible here. The short time memory is visible as a blue shadow, which for highly stimulated part (the middle stroke of digit "4", Figure 14 b) extends behind the edge of the image (so the observable practical memory capacity is at least 14 time steps, further evaluation is not possible due to structure of the data set). These effects: neuron-like hyperpolarization and short time memory contribute to the computing performance of the device.

For evaluating the voice recognition task, the Free Spoken Digit Dataset (FSDD) was used which contains six different people pronouncing ten different digits fifty times [111].

Each voice was converted to voltage versus time series signals and then applied a feature extraction algorithm called Mel-Frequency Cepstral Coefficient (MFCC) [112, 113]. The reason for using a feature extraction algorithm is to simplify the high dynamic voice to make it easier to classify signals. In the experiment part, the MFCC feature extracted signal was fed to one pad of the 16-electrode material reservoir device and 15 outputs were collected simultaneously. This process is repeated for all voice signals for training. In the classification part, there are two different tasks performed one is one person pronouncing ten different digits, other one is six people pronouncing the same digit. Each MFCC voice digit is a vector with 200 raw long, with 6 people- named as George (G), Jackson(J), Lucas (L), Nicolas (N), Theo (T), Yeweweler (Y). For each person, 45 of 50 sounds were trained with a feed-forward line with a back-propagation algorithm. The obtained weight values were tested on the remaining 5 out of 50 sounds to determine the classification accuracy of the reservoir device. The Accuracy percentage after 2000 epoch training gets constant at ~ 84% (Figure 15).

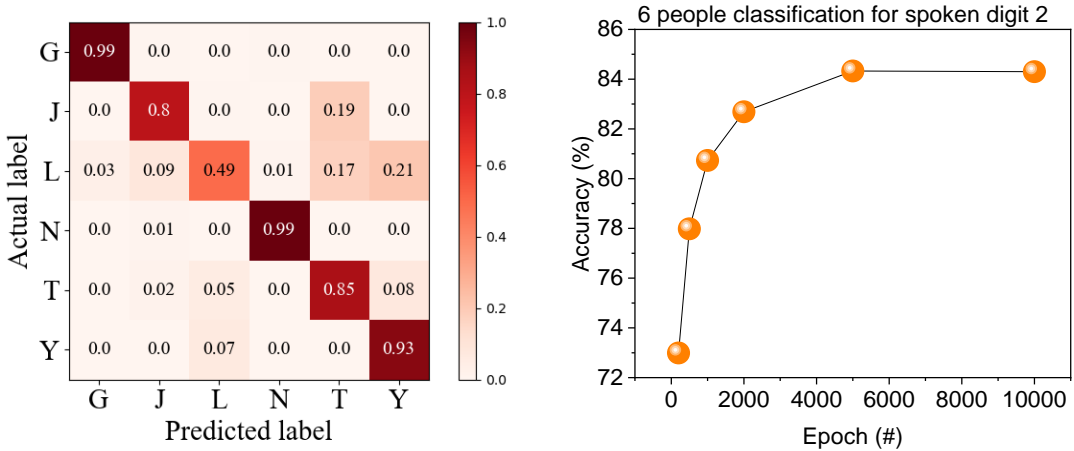

Figure 15. Confusion matrix for digit 2 for six different people after 1000 epoch. Accuracy plot after applying different epoch for training of output from reservoir.

Table 2 presents a performance comparison of MNIST classification utilizing physical reservoir computing (PRC) with other neural network-based devices employing different kinds of materials. The proposed devices exhibit diverse capability in handling the classification task while achieving higher accuracy with reduced training overhead is achievable with PRC. This advantage is primarily due to the compact network architecture, which enhances efficiency.

**Table 2.** The benchmarking of MNIST classification task using not only PRC but also other neural network-based devices based on different materials or simulated systems.

| System | Accuracy percentage | | |
|---|---|---|---|
| Trained linear perceptron | 74 | CNN | [133] |
| Simulated reservoir | 99.25 | RC+ CNN | [141] |
| Al/PMMA/BiOBr/FTO | 94.01 | CNN | [134] |
| Au/$Bi_2Ti_2O_7$/ITO | 97.7 | CNN | [135] |
| $MoS_2$/$Al_2O_3$/ $Hf_{0.5}Zr_{0.5}O_2$/W transistor | 95.43 | PRC | [136] |
| Ti/$TiO_x$/$TaO_y$/Pt[a] | 98.44 | RC | [137] |

| Pt/Ti/SiO$_2$/Ag/Pd | 83 | PRC | [138] |
| Au/[emim$^+$][TFSI$^-$]/Au | 87.3 | PRC | [139] |
| Ag/hBN/GN/PEDOT:PSS/Ag$^a$ | 88.1 | RC | [140] |
| Au/SiO$_2$/CoO$_x$/ITO | 93.39 $^b$ | PRC | [142] |
| Ag/(BA)$_2$MA$_3$Pb$_4$I$_{13}$/ITO | 92.17 | PRC | [143] |
| Pt/GDC/CeO$_2$/Pt | 90.5 | PRC | [144] |
| ZnO/Ta$_2$O$_5$SE-FET | 95.41 | PRC | [145] |
| W/HfO$_2$/TiN | 90.1$^c$ | PRC | [146] |
| Pd/Au/WO$_x$/W | 88.1 | PRC | [147] |
| OESTs | 94.9 | PRC | [148] |
| Ag/PMMA/Cs$_3$Cu$_2$I$_5$/ITO | 94 | MLP | [125] |
| Ag/PMMA/Cs$_2$AgBiBr$_6$/ITO | 91 | three-layer neural network | [128] |
| Ag/Col NFs/FTO | 83.85 | SNN | [149] |
| Au/7-MeqBiI$_3$/Au | 97 | PRC | This work |

$^a$ Simulated; $^b$ software readout mode result in 96.89 % accuracy; $^c$ adding a hidden layer result in 96.5% accuracy. Abbriviations: CNN: convolutional neural network; [emim$^+$]: 1-ethyl-3-methylimidazolium; [TFSI$^-$]: trifluoromethane sulfonyl)imide; hBN: hexagonal boron nitride; GN: graphene; BA: butylammonium; SE-FET: nano-ionic solid electrolyte field effect transistor; GDC: Pt/Gd-doped CeO$_2$; OESTs: organic electrochemical synaptic transistors; Col: collagen; NFs: 1D nanofibers; SE-FET: nano-ionic solid electrolyte field effect transistor; FA: formamidinium cation; MLP: multilayer perceptron; SNN: spiking neural network.

## 4. Conclusion

This study examines the effect of halide ligand on the structure and semiconducting properties of 7-methylquinolinium halobismuthates. The variation in halogen size influences structural organization. For example, the smaller Cl ligands allow for more compact and interconnected zigzag chains, while larger I form less tightly packed chains or discrete motifs. The organic cation 7-methylquinoline is common across all structures, but its arrangement varies. Moreover the disorder in organic cations highlights flexibility in organic motifs, which might contribute to dynamic properties under external stimuli. The organic and inorganic components are connected via hydrogen bonding (e.g., N-H⋯X interactions) and weaker interactions (C-H⋯X). These interactions stabilize the crystal lattice and contribute to structural diversity.

Notably, interactions like N-H⋯Br in 7-MeqBiBr$_3$ and N-H⋯Cl in 7-MeqBiCl$_3$ are stronger and more directional than weaker C-H⋯X interactions, leading to different packing efficiencies. Overall, the size of the halogen affects not only the inorganic motif (0D or 1D) but also the nature and strength of intermolecular interactions, influencing the overall lattice stability and electronic properties. This is reflected in the dramatical changes in optical band gaps from 2.1 in iodide to 3.1 in chloride derivatives, measured by DRS technique and confirmed through DFT calculations.

Sandwiched thin film between ITO and Cu of 7-MetqBiI$_3$, demonstrated memristive behavior with pinched hysteresis loops in I-V scans. Conductivity studies under varying temperatures revealed activation energies less than ~ 0.3 eV. Its performance is comparable with other coordination compounds-based memristive devices, it is characterized by relatively high retention time as well as switching endurance (Table S8). Its low ON/OFF ratio is a consequence of interfacial switching as well as relatively high conductivity of the materials themselves.

Therefore it is more suitable for neuromorphic computing applications (e.g .reservoir computing) rather than memory applications. Neuromorphic properties, including nonlinear potentiation and depression, showed potential for physical reservoir computing. Benchmark tasks tested complexity and non-linearity of devices with 16 electrodes (one input and 15 outputs), achieving 82.26% accuracy in MNIST classification and 82% in voice recognition tasks. These findings highlight the suitability of these materials for advanced computing applications and indicate how simple physical reservoir approach, supplemented with linear perceptron provides high computational performance, superior to other related systems.

**Acknowledgments**

The authors acknowledge the financial support from the Polish National Science Centre within the OPUS programme (grant agreement no. UMO-2020/37/B/ST5/00663) and AGH University of Science and Technology within the program "Excellence Initiative-Research University". The authors gratefully acknowledge Polish high-performance computing infrastructure PLGrid (HPC Center: ACK Cyfronet AGH) for providing computer facilities and support within computational grant no. PLG/2024/017405.

**Declaration of competing interest**

All the contributing authors report no conflict of interests in this work.

**Author contribution statement**

Author 1: Data curation, Investigation, writing manuscript, experimental design. Author 2: experimental design. Author 3: experimental design. Author 4: experimental design, writing manuscript. Author 5: experimental design. Author 6: experimental design. Author 7: experimental design. Author 8: Data curation, Investigation, project administration, funding acquisition, validation, writing manuscript, experimental design. All the authors have approved the final manuscript.

# Supporting Information

Table S1. X-ray diffraction data and refinement details

|  | 7-MeqBiI$_3$ | 7-MeqBiI$_3$-I | 7-MeqBiBr$_3$ | 7-MeqBiCl$_3$ |
|---|---|---|---|---|
| Chemical formula | C$_{10}$H$_{10}$Bi$_{0.50}$I$_{2.50}$N | C$_{10}$H$_{10}$BiI$_4$N | C$_{20}$H$_{20}$BiBr$_5$N$_2$ | C$_{10}$H$_{10}$Bi$_{0.50}$Cl$_{2.50}$N |
| M$_r$ | 565.93 | 860.77 | 896.91 | 337.30 |
| Crystal system, space group | Monoclinic, C2/m | Triclinic, P-1 | Triclinic, P-1 | Monoclinic, C2/c |
| Lattice parameters (Å)/° | a = 13.9025(2)<br>b = 20.1091(2)<br>c = 9.9870(1)<br>b= 107.754(1) | a = 7.7115(2)<br>b = 11.0571(2)<br>c = 11.0627(3)<br>a= 82.465(2)<br>b= 72.143(2)<br>g = 72.191(2) | a = 10.4835(2)<br>b = 10.8664(3)<br>c = 12.1337(3)<br>a= 106.929(2)<br>b= 107.879(2)<br>g = 99.839(2) | a = 18.5917(3)<br>b = 16.6441(3)<br>c = 7.3962(1)<br>b= 97.781(1) |
| V (Å$^3$) | 2659.06(6) | 854.09(4) | 1205.97(5) | 2267.62(6) |
| Z | 8 | 2 | 2 | 8 |
| D$_x$(g/cm$^3$) | 2.827 | 3.347 | 2.467 | 1.976 |
| T (K) | 100(2) K | 100(2) | 130(2) | 100(2) |
| Radiation type | MoKa | MoKa | MoKa | MoKa |
| μ (mm$^{-1}$) | 12.445 | 17.530 | 15.603 | 8.374 |
| Theta range (°) | 2.948 to 33.648 | 2.681 to 33.496 | 2.130 to 33.355 | 3.101 to 33.624 |
| Absorption correction | Multi-scan | Gaussian | Gaussian | Multi-scan |
| Crystal size (mm) | 0.30x0.25x0.12 | 0.29x0.09x0.06 | 0.39x0.23x0.09 | 0.25x0.22x0.10 |
| Data/restraints/parameters | 5089 / 22 / 235 | 5894 / 184 / 248 | 8175 / 0 / 256 | 4247 / 0 / 134 |
| R(int) | 0.1006 | 0.0503 | 0.0643 | 0.0738 |
| Goodness-of-fit | 1.106 | 1.024 | 1.018 | 1.096 |
| Final R indices [I>2sigma(I)] | R1 = 0.0272, wR2 = 0.0643 | R1 = 0.0291, wR2 = 0.0661 | R1 = 0.0347, wR2 = 0.0739 | R1 = 0.0197, wR2 = 0.0459 |
| Δρ$_{min}$/Δρ$_{max}$ (e/Å$^3$) | 1.870 and -1.911 | 2.859 and -2.130 | 2.765 and -2.106 | 1.784 and -1.501 |
| Packing coefficient* | 0.735 | 0.716 | 0.694 | 0.698 |

*CSD average packing coefficient for organometallic molecules 0.67(5)

CCDC Deposit numbers: 2330350, 2330351, 2330352, 2330353

Table S2 Bi-I bond lengths in the examined structures.

| 7-MeqBiI$_3$ | | 7-MeqBiBr$_3$ | [Å] |
| --- | --- | --- | --- |
| Bi(1)-I(4) | 2.9300(4) | Bi(1)-Br(6) | 2.7105(4) |
| Bi(1)-I(5) | 2.9815(4) | Bi(1)-Br(4) | 2.7496(4) |
| Bi(1)-I(3)#1 | 3.0792(2) | Bi(1)-Br(3) | 2.7619(4) |
| Bi(1)-I(3) | 3.0792(2) | Bi(1)-Br(5) | 2.9698(4) |
| Bi(1)-I(2)#2 | 3.1916(3) | Bi(1)-Br(2) | 3.0213(4) |
| Bi(1)-I(2) | 3.3140(4) | Bi(1)-Br(2)#1 | 3.1022(4) |
| #1 x,-y+1,z | | Br(2)-Bi(1)#1 | 3.1023(4) |
| #2 -x+1,-y+1,-z | | | |
| | | #1 -x+1,-y+1,-z+1 | |
| 7-MeqBiI$_3$-I | [Å] | 7-MeqBiCl$_3$ | [Å] |
| Bi(1)-I(3) | 2.9097(3) | Bi(1)-Cl(1)#1 | 2.5606(4) |
| Bi(1)-I(5) | 2.9101(3) | Bi(1)-Cl(1) | 2.5606(4) |
| Bi(1)-I(4) | 3.0874(3) | Bi(1)-Cl(2)#1 | 2.7016(5) |
| Bi(1)-I(2) | 3.0875(3) | Bi(1)-Cl(2) | 2.7016(4) |
| Bi(1)-I(2)#1 | 3.2996(3) | Bi(1)-Cl(3) | 2.8856(1) |
| Bi(1)-I(4)#2 | 3.3004(3) | Bi(1)-Cl(3)#1 | 2.8856(1) |
| #1 -x+1,-y+1,-z+1 | | #1 1-x, y, ½-z | |
| #2 -x+2,-y+1,-z+1 | | | |

Table S3 Geometry of weak interactions in 7-MeqBiI$_3$ [Å/°]

| D-H···A | d(D-H) | d(H···A) | d(D···A) | <(DHA) |
|---|---|---|---|---|
| C(9^a)-H(9^a)...I(2)#3 | 0.95 | 3.10 | 4.039(9) | 170.3 |
| N(1^a)-H(1^a)...I(3)#4 | 0.88 | 2.76 | 3.510(6) | 144.4 |
| C(2^a)-H(2^a)...I(4)#5 | 0.95 | 3.21 | 4.097(8) | 155.2 |
| C(2A^b)-H(13^b)...I(4)#6 | 0.95 | 3.09 | 3.986(12) | 158.2 |
| C(1A^b)-H(12A^b)...I(2)#3 | 0.95 | 3.31 | 3.813(14) | 115.2 |
| C(9A^b)-H(19^b)...I(3)#4 | 0.95 | 3.29 | 4.190(13) | 158.2 |
| Symmetry transformations used to generate equivalent atoms:<br>#1 x,-y+1,z   #2 -x+1,-y+1,-z   #3 -x+3/2,-y+3/2,-z+1<br>#4 x+1/2,-y+3/2,z+1   #5 -x+2,-y+1,-z+1   #6 -x+3/2,-y+3/2,-z ||||| 

Table S4 Geometry of weak interactions in 7-MeqBiBr$_3$ [Å/°]

| D-H···A | d(D-H) | d(H···A) | d(D···A) | <(DHA) |
|---|---|---|---|---|
| N(1A)-H(1A)...Br(2)#2 | 0.88 | 2.48 | 3.339(3) | 164.4 |
| C(8A)-H(8A)...Br(2)#2 | 0.95 | 3.14 | 3.878(4) | 136.2 |
| N(1B)-H(1B)...Br(5) | 0.88 | 2.44 | 3.265(3) | 157.1 |
| C(2A)-H(2A)...Br(6)#3 | 0.95 | 2.87 | 3.767(4) | 158.1 |
| C(1A)-H(1A1)...Br(4)#2 | 0.95 | 2.82 | 3.714(4) | 156.2 |
| C(8B)-H(8B)...Br(2) | 0.95 | 3.09 | 3.996(4) | 161.1 |
| Symmetry transformations used to generate equivalent atoms:<br>#1 -x+1,-y+1,-z+1   #2 -x+1,-y+2,-z+1   #3 x+1,y+1,z ||||| 

Table S5 Geometry of weak interactions in 7-MeqBiCl$_3$ [Å/°]

| D-H···A | d(D-H) | d(H···A) | d(D···A) | <(DHA) |
|---|---|---|---|---|
| N(1)-H(11)...Cl(2)#3 | 0.87(3) | 2.28(3) | 3.1452(18) | 172(2) |
| C(1)-H(1)...Cl(3) | 0.95 | 2.91 | 3.716(2) | 143.0 |
| C(2)-H(2)...Cl(1) | 0.95 | 2.96 | 3.610(2) | 127.1 |
| C(9)-H(9)...Cl(2)#3 | 0.95 | 2.75 | 3.5296(19) | 139.4 |

Symmetry transformations used to generate equivalent atoms:
#1 -x+1,y,-z+1/2   #2 -x+1,-y+1,-z+1   #3 x,-y+1,z+1/2

Table S6 Geometry of π···π interactions in 7-MeqBiBr$_3$ [Å/°]

| Cg···Cg | Distance [Å] |
|---|---|
| Cg1···Cg6 | 3.598 |
| Cg1···Cg2 | 3.517 |
| Cg2···Cg4 | 3.580 |
| Cg5···Cg3 | 3.580 |
| Cg1: C4B, C5B, C6B, C7B, C8B, C9B | Cg4: N1A$^i$, C1A$^i$, C2A$^i$, C3A$^i$, C4A$^i$, C9A$^i$ |
| Cg2: C4A, C5A, C6A, C7A, C8A, C9A | Cg5: N1A$^{ii}$, C1A$^{ii}$, C2A$^{ii}$, C3A$^{ii}$, C4A$^{ii}$, C9A$^{ii}$ |
| Cg3: N1A, C1A, C2A, C3A, C4A, C9A | Cg6: C4B$^{iii}$, C5B$^{iii}$, C6B$^{iii}$, C7B$^{iii}$, C8B$^{iii}$, C9B$^{iii}$ |
| | i: 1-x, 2-y, -z |
| | ii: 1-x, 2-y, -z |
| | iii: 2-x, 2-y, 1-z |

Table S7. Switching characteristics for small (1 mm$^2$) and large (9 mm$^2$) electrodes determined for the ITO/7-MeqI$_3$:/Cu devices.

| Surface area: 1 mm$^2$ | | Surface area: 9 mm$^2$ | | ON/ON ratio (9 mm$^2$ vs 1 mm$^2$) | OFF/OFF ratio (9 mm$^2$ vs 1 mm$^2$) |
|---|---|---|---|---|---|
| ON current (mA) | OFF current (mA) | ON current (mA) | OFF current (mA) | | |
| 5.18±0.4 | 2.25±0.3 | 7.97±0.4 | 3.07±0.2 | 1.5 | 1.36 |
| ON/OFF ratio: 2.30 | | ON/OFF ratio: 2.60 | | | |

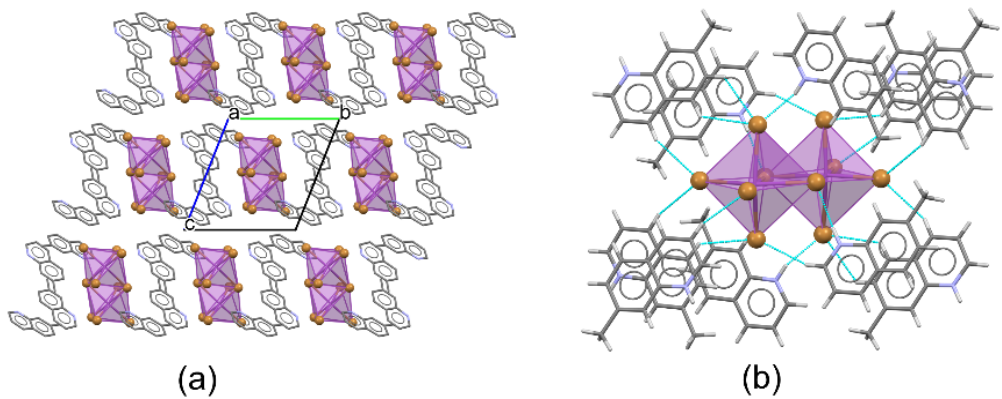

Figure S1. a) and c) packing of the structural components viewed along [100] crystallographic direction in the 7-MeqBiBr$_3$; b) and d) interactions between [Bi$_2$X$_{10}$] units and organic cations in the structures of 7-MeqBiBr$_3$.

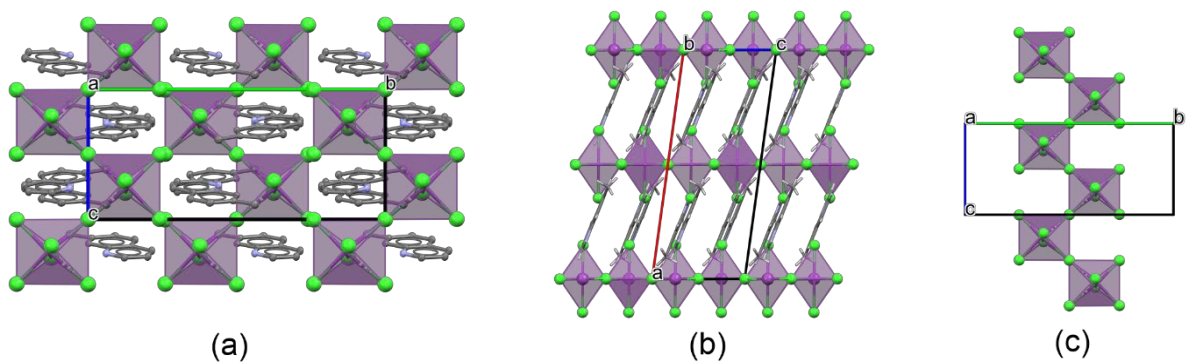

Figure S2. 7-MeqBiCl$_3$: a) packing of the structural components viewed along [100] direction; b) packing of the structural components viewed along [010] direction and c) 1D zig-zag chain.

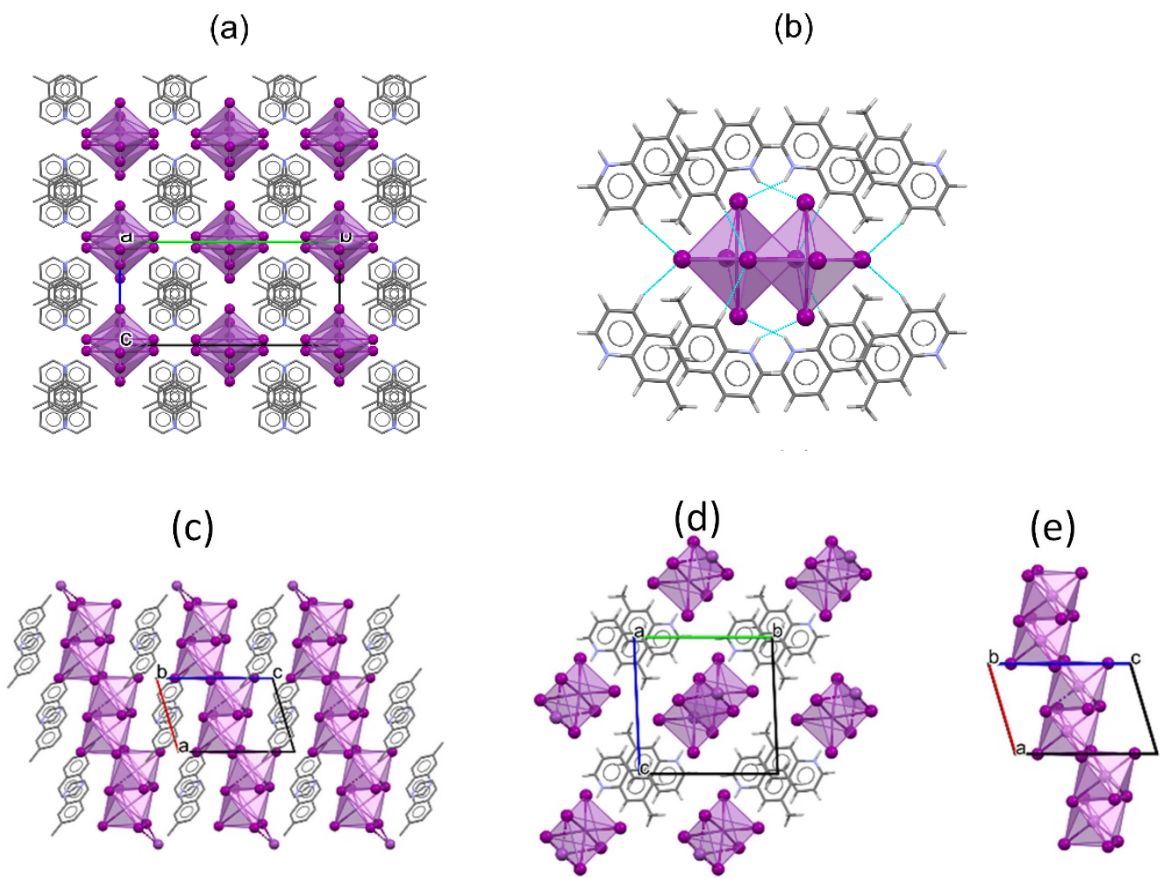

Figure S3. 7-MeqBiI$_3$: a) packing of the structural components viewed along [100] crystallographic direction, b) interactions between [Bi$_2$X$_{10}$] units and organic cations. 7-MeqBiI$_3$-I: c) packing of the structural components viewed along [010] direction, d) packing of the structural components viewed along [100] direction and e) 1D-chain.

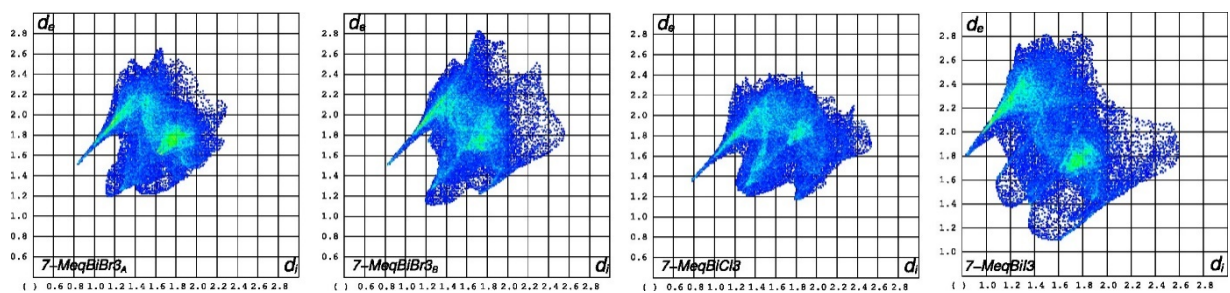

Figure S4. Fingerprint plots for the crystal structures of 7-MeqBiX$_3$ (X: -I, -Cl, -Br).

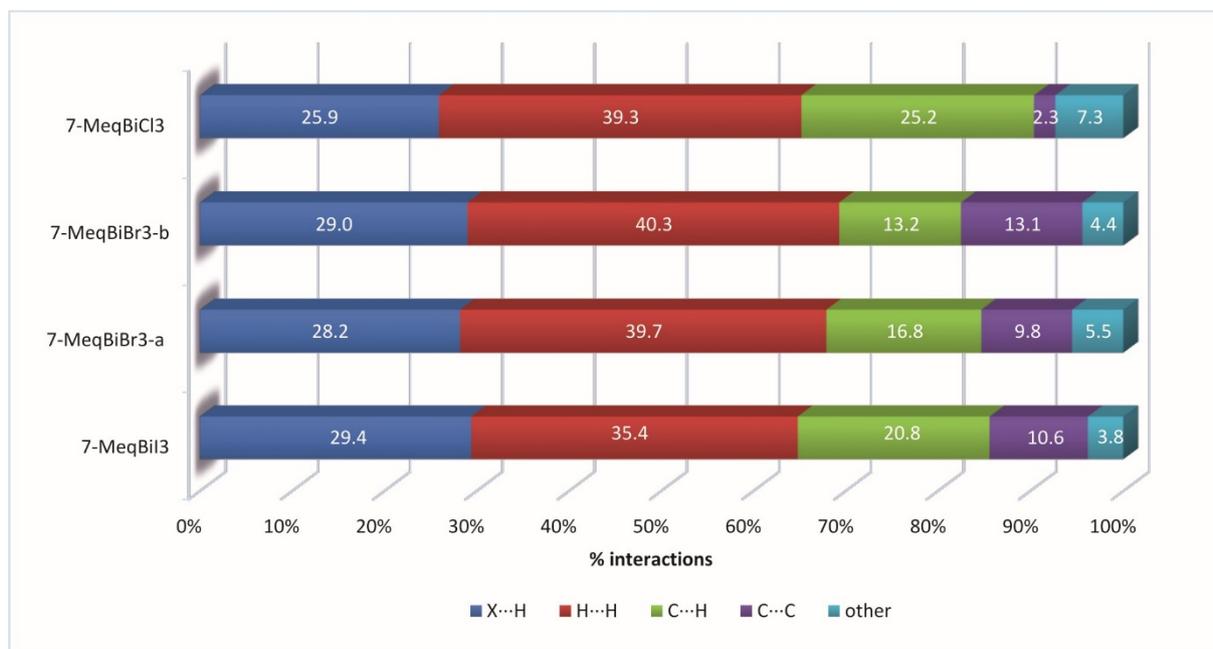

Figure S5. Percentage contribution of each interactions to the Hirshfeld surface.

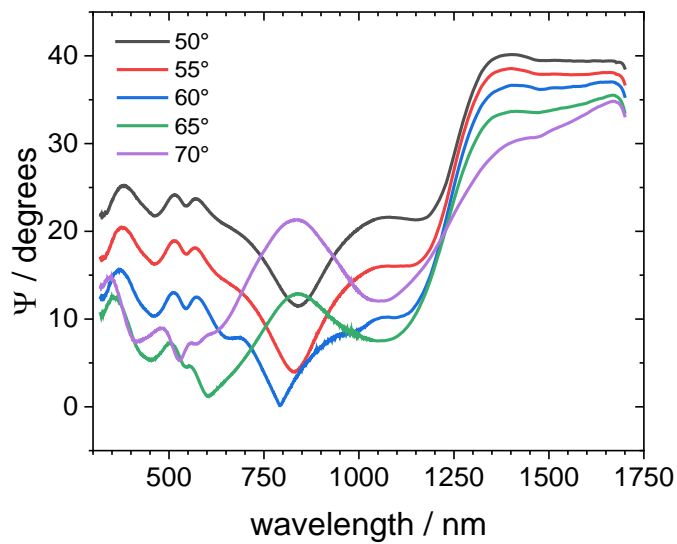

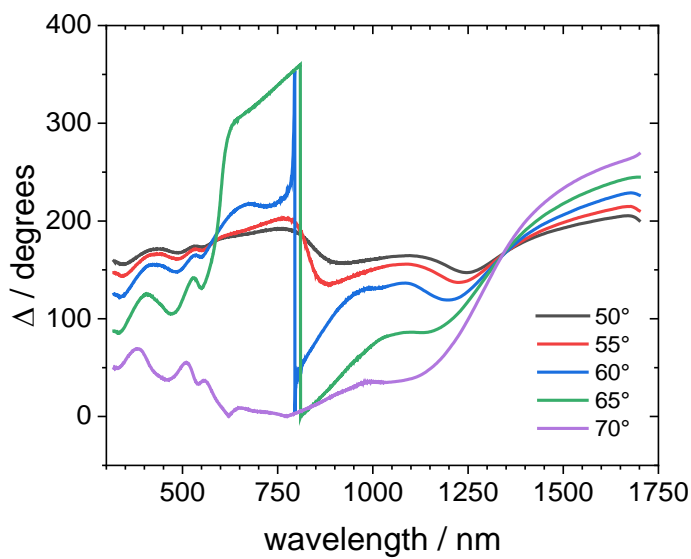

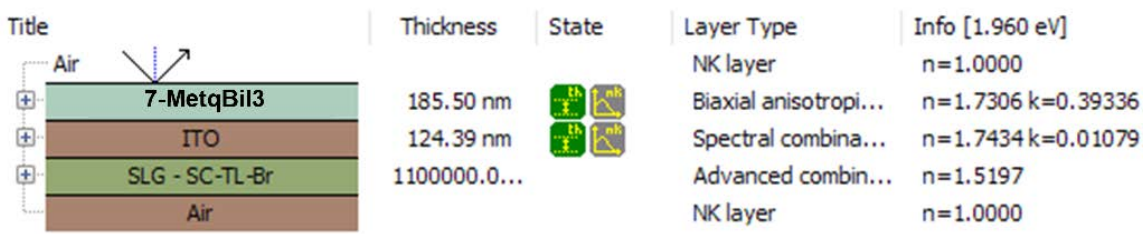

Figure S6. Experimental ellipsometry angles Ψ, Δ and the fitting theoretical model for this layers of 7-MetqBiI₃ aon ITO glass.

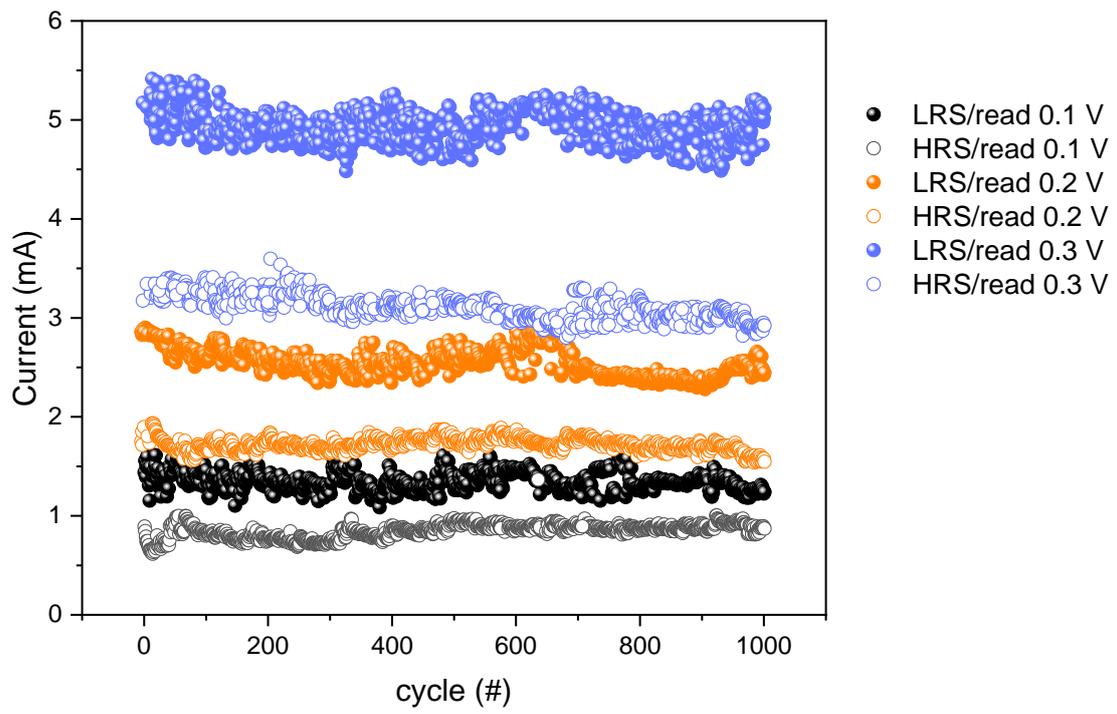

Figure S7. ON-OFF measurements with different read voltage as 0.1 V, 0.2 V, and 0.3 V.

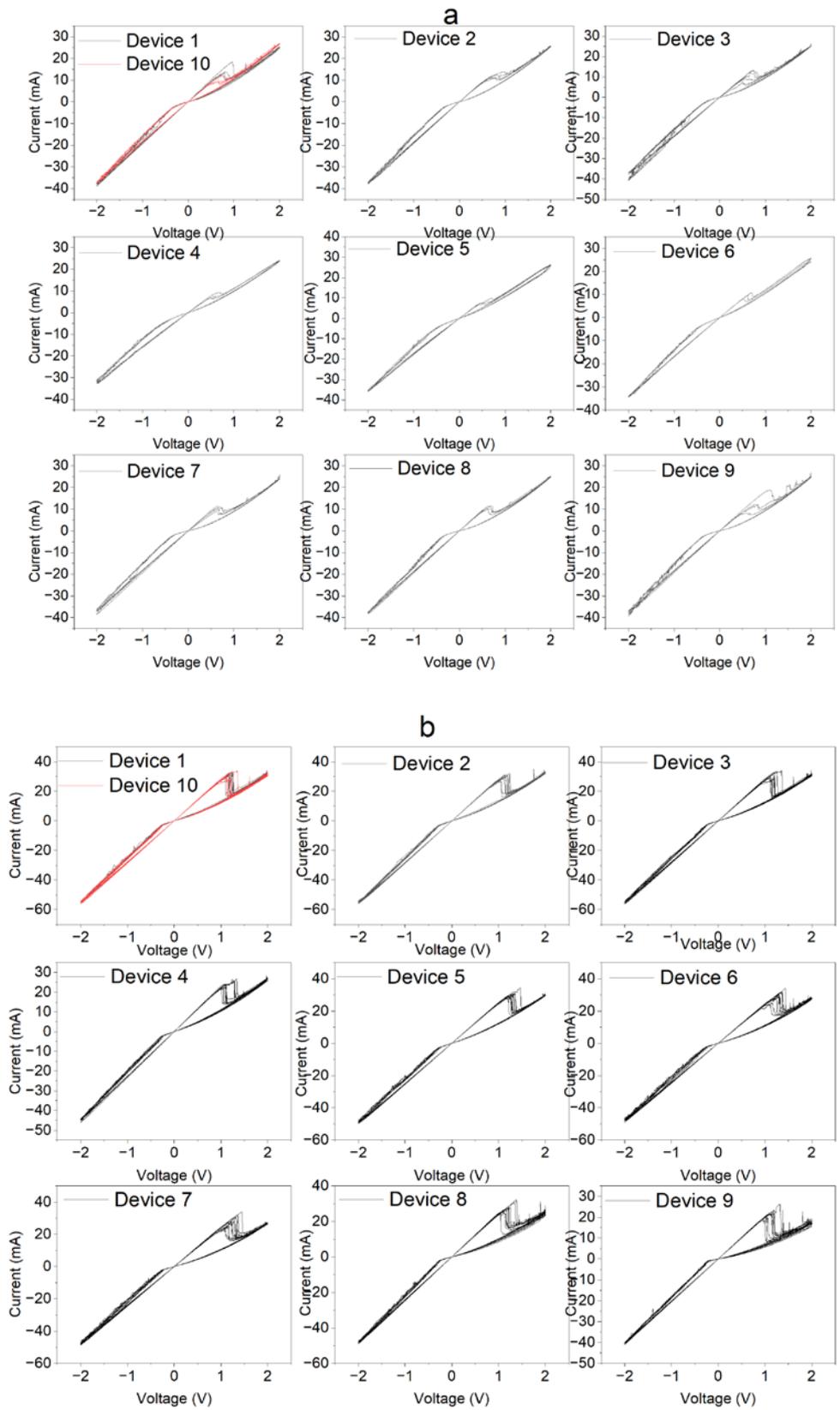

Figure S8. IV plots for devices with two different sizes of top electrodes with surface area of: a) 1mm$^2$ b) 9 mm$^2$

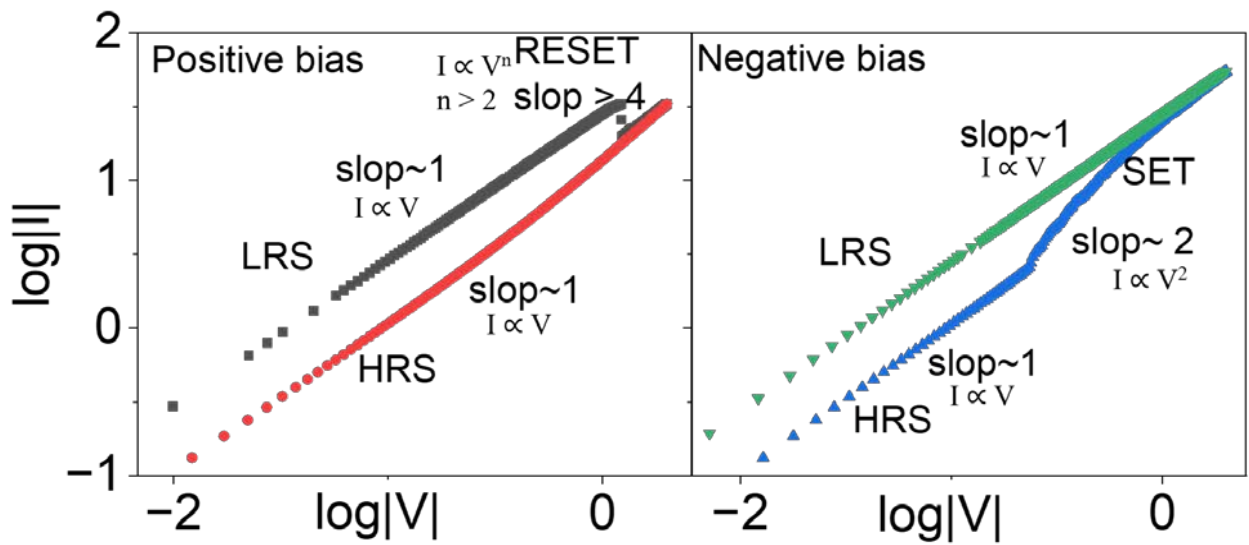

Figure S9. Double-logarithmic plots of current-voltage (I-V) characteristics for Cu/7-MeqBiI$_3$/ITO.

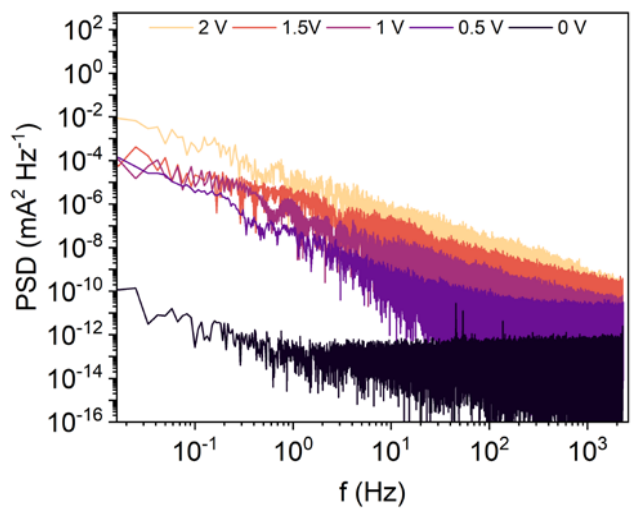

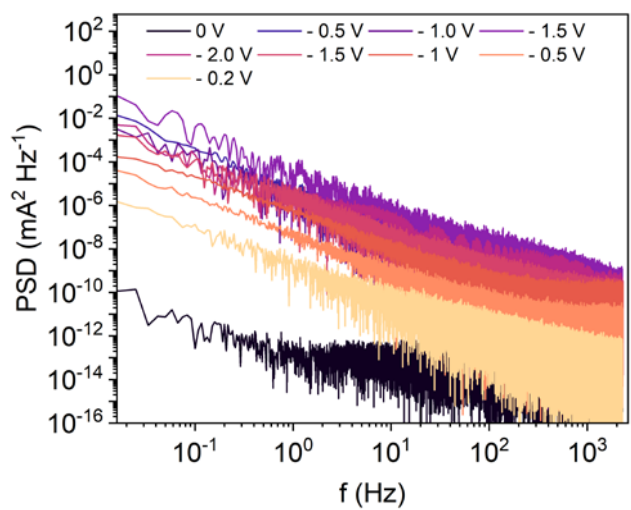

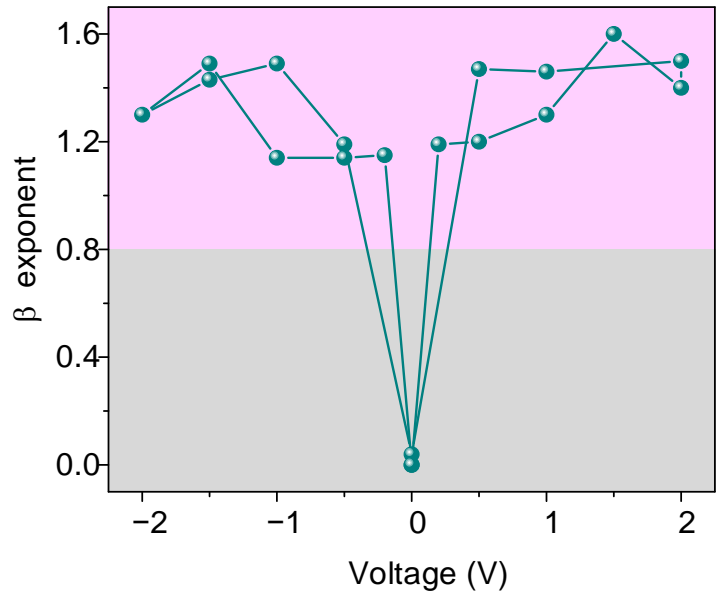

Figure S10. Noise analysis of Cu/7-MeqBiI$_3$/ITO/glass at a range of applied forward and reverse biased for (+2 V to -2 V).

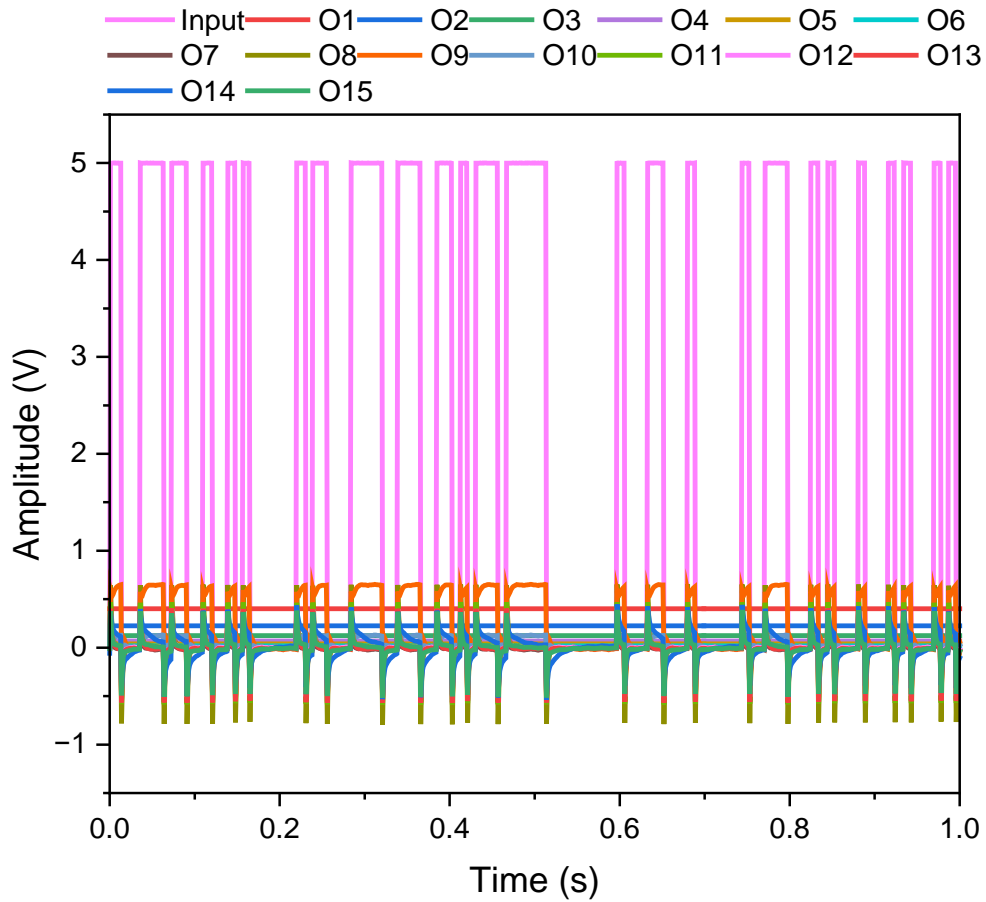

Figure S11. Applied input pulse sequences for memory capacity evaluation over 1 s with 1000 point per second sampling rate and fifteen outputs.
49